\documentclass[reprint,amsmath,amssymb,aps,prx,eqsecnum,longbibliography,superscriptaddress]{revtex4-2}
\usepackage[usenames,dvipsnames]{color}
\usepackage[colorlinks=true,linkcolor=Blue,citecolor=Blue,urlcolor=Blue]{hyperref}
\hypersetup{
  pdftitle={Landau theory of quenched criticality in linear in-context learning},
  pdfauthor={Seungho Lee, Hyojae Jeon, and Jung Hoon Han},
  pdfsubject={Statistical physics and Landau theory of linear in-context learning}
}
\usepackage{graphicx}
\usepackage{verbatim}
\usepackage{bbm}
\usepackage{bm}
\usepackage{amsmath} 
\usepackage{amssymb}
\usepackage{mathptmx}
\usepackage[version=3]{mhchem}
\usepackage{float}
\usepackage{subcaption}
\usepackage{kotex}
\usepackage{booktabs}
\usepackage{multirow}
\graphicspath{{figures/}}

\usepackage{caption}
\newcommand{\ba}{\begin{eqnarray}}
\newcommand{\ea}{\end{eqnarray}}
\newcommand{\bd}{\begin{displaymath}}

\newcommand{\nn}{\nonumber \\}

\newcommand{\R}{\mathbb{R}}

\newcommand{\x}{{\bf x}}

\newcommand{\dd}{{\rm d}}

\begin{document}
\title{Landau theory of quenched criticality in linear in-context learning}
\author{Daesik \surname{Kim}}
\affiliation{Department of Physics, Sungkyunkwan University, Suwon 16419, South Korea}

\author{Sumin \surname{Choi}}
\affiliation{Department of Physics, Sungkyunkwan University, Suwon 16419, South Korea}

\author{Hyojae \surname{Jeon}}
\affiliation{Department of Physics, Sungkyunkwan University, Suwon 16419, South Korea}

\author{Jung Hoon \surname{Han}}
\email{hanjemme@gmail.com}
\affiliation{Department of Physics, Sungkyunkwan University, Suwon 16419, South Korea}
\affiliation{Center for Quantum Dynamics of Angular Momentum, Pohang University of Science and Technology, Pohang 37673, Korea}

\begin{abstract}  
In-context learning (ICL) allows a pretrained model to infer a new task from examples supplied in its prompt without updating its parameters. In linear models of ICL, the prediction error develops a double-descent singularity when the number of pretraining samples becomes comparable to the number of learnable parameters. We formulate this interpolation singularity as a critical phenomenon of a quenched disordered system. By comparing annealed and quenched descriptions of the same linear ICL model, we identify the connected sample-to-sample fluctuations of the learned parameters as the microscopic origin of the singular error. A Landau potential is constructed by integrating the cavity self-consistency equation for the renormalized ridge parameter $\xi$. The role of (magnetization) order parameter is played by $\xi$, while the bare ridge parameter $\lambda$ becomes its conjugate magnetic field. The normalized sample complexity $\tau$ acts as a temperature and the double-descent singularity occurs at the critical temperature $\tau_c =1$. The Landau susceptibility is precisely the quantity that diverges in the fluctuation contribution to the prediction error. The order parameter is closely related to the fraction of zero eigenvalues of the empirical relaxation matrix in the ridgeless limit, which define flat directions in the learning dynamics. The Landau theory is generically cubic in the order parameter with critical exponents $(\beta_{\rm cr},\delta_{\rm cr},\gamma_{\rm cr})=(1,2,1)$. In the large-context regime, there appears a pseudogap-like regime characterized by suppressed order parameter. Predictions of the Landau theory are independently confirmed from numerical solutions of the original learning problem with good quantitative agreement. Our results pave the way for solid statistical-physics understanding of the interpolation criticality in linear in-context learning.
\end{abstract}
\date{\today}
\maketitle

\section{Introduction}

A regression problem seeks to infer the relation underlying a set of input--output examples and to use that information to predict the output associated with a new input. Consider, for example, a family of linear tasks in which an unknown task vector $\mathbf{w}$ and random noises $\sigma\epsilon_m$ generate the output
\begin{equation}
    y_m=\mathbf{w}\cdot\mathbf{x}_m+\sigma\epsilon_m .
\end{equation}
Given a sequence of demonstrations $\{(\mathbf{x}_m,y_m)\}_{m=1}^{\ell}$ and a new query
$\mathbf{x}_{\ell+1}$, the objective is to predict the corresponding $y_{\ell+1}$. In conventional regression, one would try to come up with a new estimator of $\mathbf{w}$ for each fresh data set. In in-context learning (ICL), once the pretraining is complete, the parameters of the network remain fixed rather than adapted to new tasks. A new set of demonstrations is supplied through the prompt, and the task-dependent prediction is generated by the pretrained network. From the physics perspective, ICL is a scheme that learns the proper ``response function" of the physical system at hand through training off of existing datasets, and gives out correct responses for fresh inputs without having to re-calibrate the response function.

The modern study of ICL was catalyzed by the observation of Brown \textit{et al.}\ that sufficiently large pretrained language models can perform new tasks from instructions and a small number of demonstrations provided through the prompt, without further parameter
updates \cite{brown20}. Subsequently, Xie \textit{et al.}\ interpreted ICL as a form of implicit Bayesian inference in which a latent task or concept is inferred from the
examples appearing in the context \cite{xie22}, while Chan \textit{et al.}\ showed that its emergence depends sensitively on the statistical structure of the pretraining distribution \cite{chan22}. Garg \textit{et al.}\ showed that Transformers trained on families of simple functions can learn previously unseen linear functions from in-context examples with performance comparable to standard regression estimators \cite{garg22}. This connection was sharpened by Aky\"urek \textit{et al.}, who demonstrated that Transformers can implement or approximate familiar learning algorithms in their forward computation \cite{akyurek23}, and by von Oswald \textit{et al.}, who established an explicit correspondence between linear self-attention and gradient-descent updates on regression problems \cite{vonoswald23}. 

The linear ICL model central to our study has the pretraining problem that is itself a high-dimensional regression; the predicted output takes
the form
\begin{equation}
    \hat y=\mathbf{P}\cdot\mathbf{H},
\end{equation}
where $\mathbf{H}$ encodes the training context and query, while the learning vector $\mathbf{P}$ contains $d(d+1)=O(d^2)$ trainable parameters. Pretraining on $n$ independent contexts amounts to fitting $\mathbf{P}$ to $n$ regression constraints. The normalized sample complexity
\begin{equation}
    \tau=\frac{n}{d^2}
\end{equation}
is the asymptotic sample-to-parameter ratio. The point $\tau=1$ is then recognized as the interpolation threshold at which the number of training constraints becomes
comparable to the number of trainable directions. 

An asymptotically precise theory of the linear ICL (LICL) model was developed by Lu \textit{et al.}\ \cite{pehlevan25}, in terms of three normalized parameters: the context length $\alpha=\ell/d$, the task diversity $\kappa=k/d$, and the sample complexity $\tau=n/d^2$. Using the cavity method, they obtained the learned predictor and the corresponding in-context learning and in-distribution generalization errors. In the ridgeless limit, the prediction error develops a double-descent singularity at the
interpolation threshold $\tau=1$. Subsequent work has investigated the role of pretrain--test task alignment and the broader scaling structure of in-context regression \cite{pehlevan25b,pehlevan25c}, while related statistical-mechanical approaches have uncovered spin-glass structure in simplified models of ICL \cite{huang25}. These developments provide increasingly precise descriptions of learning curves and generalization, but still leave open an elementary physics question: {\it what quantity fluctuates critically at the interpolation threshold, and can the singularity be understood as a critical phenomenon in the spirit of Landau theory}?

The question may have broader implication outside the scope of linear ICL. It has been shown that a qualitative change in statistical behavior can occur when the number of constraints becomes comparable to the number of trainable degrees of freedom~\cite{belkin19,hastie22,bartlett20,nakkiran21}. Random-matrix theory further shows that the resulting prediction risk is controlled by the low-lying spectrum of empirical covariance and Gram matrices \cite{dobriban18}. From the perspective of statistical physics, learning problems with random training sets are instances of disordered systems where decades of extensive literature exist. The distinction between annealed and quenched averages, the connected correlations that survive quenched disorder averaging, and their relation to susceptibilities and phase transitions are all central themes in the statistical mechanics of disordered systems~\cite{seung92,zdeborova16,bahri20,opper01}. Yet a firm connection between the physics of disordered systems and the observed interpolation peak in the theory of learning has not been established.

We pursue this viewpoint to its logical conclusion by developing an appropriate Landau theory of linear ICL near the threshold. Firstly, an annealing approach is developed to calculate the learning vector ${\bf P}$ and its results are compared to those of the cavity method of \cite{pehlevan25}, which we now dub the quenching scheme. By comparing the two approaches, we firmly identify the divergence of training sample-to-sample fluctuations of the learning parameters to be the culprit behind the observed double descent behavior of the prediction error. 

The cavity solution contains a self-consistent renormalized ridge parameter $\xi$. A central observation of this work is that its self-consistency equation can be integrated to define a Landau potential $V(\xi)$. The normalized sample complexity $\tau$ then appears as a temperature-like control parameter, while the ridge regularization $\lambda$ enters as the conjugate field $h=\tau\lambda$. The curvature $V''(\xi_*)$ at the stationary point defines a susceptibility
\begin{equation}
\chi=\frac{1}{V''(\xi_*)} . 
\end{equation}
The exact same susceptibility appears in the singular connected-fluctuation contribution to the quenched prediction error. The double-descent divergence is thus identified with a diverging Landau susceptibility.

The Landau order parameter $\xi_*$ possesses a concrete geometric meaning. In the ridgeless limit, $\xi_*$ is related to the fraction of zero eigenvalues of the empirical relaxation matrix. Each zero eigenvalue defined a ``flat direction" along which no learning process takes place. The ordered phase is characterized by an extensive null space in the learning space rather than by conventional spontaneous symmetry breaking. The lack of symmetry-breaking interpretation also explains why the Landau functional generically contains a cubic term in $\xi$.

For every fixed finite inverse context length $\gamma=(1+\sigma^2)/\alpha>0$, the asymptotic critical point lies at $\tau_c=1$. The cubic Landau theory gives the critical exponents
\begin{equation}
(\beta_{\rm cr},\delta_{\rm cr},\gamma_{\rm cr})=(1,2,1),
\end{equation}
which obey the Widom scaling relation. The phase diagram structure becomes considerably richer when the context length is large ($\gamma\ll1$) and the task diversity $\kappa$ is varied. For $\kappa<1$, there appears a {\it crossover line} $\tau=\kappa$ separating what we call the ``pseudogap" region $\kappa < \tau < 1$ with highly suppressed order parameter $\xi_* \sim O(\gamma$). In the strict infinite-context limit $\gamma\rightarrow0$ the crossover line $\tau=\kappa$ becomes a new phase boundary. 

The endpoint $\kappa=1$, in the strict infinite-context limit, is marked by a nonanalytic term $\xi^{5/2}$ taking over the cubic term in the Landau potential. The point $(\kappa,\tau)=(1,1)$ becomes a nonanalytic critical point with exponents
\begin{equation}
(\beta_{\rm cr},\delta_{\rm cr},\gamma_{\rm cr})=\left(2, 3/2,1 \right),
\end{equation}
again satisfying Widom scaling. For small but finite $\gamma$, these exponents describe an intermediate crossover regime. 

Predictions of the Landau theory are tested directly in the original learning problem. We extract $\xi_*$ directly by fitting the analytically predicted form of the quenched learning vector to numerical solutions obtained from explicitly generated pretraining samples. The resulting $(\kappa,\tau)$ phase diagram reproduces both the strong suppression of the order parameter in the predicted pseudogap region and its crossover boundaries. The numerical results provide a strong confirmation that the Landau scheme is an excellent effective description of the linear ICL problem.

The paper is organized as follows. Section~II formulates the linear ICL problem using physics-friendly terms and notations. Section~III develops the annealed theory of ICL and shows that it contains no interpolation singularity. Section~IV reviews the quenched cavity solution and isolates the connected fluctuation contribution as the source of divergent test error. Section~V constructs the Landau functional with proper identification of the order parameter and its conjugate field. Implications of the Landau theory such as critical exponents, pseudogap regime are discussed. Section~VI tests the predictions of Landau theory numerically by extracting the order parameter directly from simulated learning problems. Section~VII summarizes the Landau-theory interpretation and shares future prospects. 

\section{Formulation of linear in-context learning} 

The theory of ICL can be mathematically phrased as the problem of predicting the output $y_{\ell+1}$ to the $d$-dimensional input ${\bf x}_{\ell+1}$ based on the known collective information regarding the input-output pairs $x_m = ( \x_m, y_m )$ ($1\le m \le \ell$) with $\x_m \in \mathbb{R}^{d}$ and $y_m \in \mathbb{R}$~\footnote{Each $\x_m$ may represent features of a house such as its location, size, number of bedrooms and bathrooms, etc. and the output $y_m$ is the house's price based on the information provided in $\x_m$.}. The italic $x_m$ is employed to denote the input-output pair as a whole; the input part alone is denoted by the bold symbol ${\bf x}_m$. The collection of data $\{( {\bf x}_1,y_1) ,\dots,( {\bf x}_\ell,y_\ell ) , {\bf x}_{\ell+1}\}$ is called the \textit{context}, and number of input-output pairs $\ell$ is called the {\it context length}. The context includes $\ell$ number of question-answer pairs, plus a fresh question ${\bf x}_{\ell+1}$ for which an answer is sought. In LICL, the estimated output $\hat y_{\ell+1}$ to a given context is governed by the equation~\cite{pehlevan25}
\begin{align}
\hat y_{\ell+1} = \frac{1}{\ell} \sum_{m=1}^\ell (q_{\ell+1} \cdot  k_m ) y_m \, . 
\label{eq:predicting-y-ell+1}
\end{align}
In a nutshell, this is a linear combination of known labels $y_m$ with weights given by the overlap between the query $q_{\ell+1} = Q x_{\ell+1}$ and the key $k_m =K x_m$ vectors for each $m$-th token vector $x_m$. There is a linear relation between existing labels $\{ y_m \}_{m=1}^\ell$ and the new one $y_{\ell+1}$ in the LICL architecture. The same formula can be written as 
\begin{align}
\hat y_{\ell+1}  = {\bf P} \cdot {\bf H} . 
\label{y-as-linear-response-main} 
\end{align}
The $d(d+1)$-dimensional context vector ${\bf H}$
\begin{align} 
{\bf H} = {\bf x}_{\ell+1}  \otimes \left( \frac{1}{\ell} \sum_{m=1}^\ell y_m \widetilde x_m \right) \, , ~~ 
\widetilde x_m \equiv
    \begin{pmatrix}
        d\,\mathbf{x}_m \\
        y_m
    \end{pmatrix} \, 
    \label{general-H} 
\end{align} 
encodes all the information contained in the context. The context information is organized in a highly nonlinear fashion through the context vector ${\bf H}$. The {\it learning vector} ${\bf P}$ determines the optimal response of the network to a given context ${\bf H}$ to produce the correct output $\hat y_{\ell+1}$. For derivation of both formulas \eqref{y-as-linear-response-main} and \eqref{general-H}\, see App.~\ref{app:how-to-derive-y-ell+1}.

The learning vector $\bf P$ is pretrained on a set of training tokens $x_m^\mu = ( \x_m^\mu , y_m^\mu )$, where $\x_m^\mu \in \mathbb{R}^{d}$ and $y_m^\mu \in \mathbb{R}$ and $1 \le m \le \ell+1$. Since this is a training set, the output value $y_{\ell+1}^\mu$ for the $(\ell+1)$-th token ${\bf x}_{\ell+1}^\mu$ is known. Training takes place over $n$ different realizations called the {\it sample complexity}, with each realization labeled by $\mu$. Readers familiar with the physics of disorder can easily associate each context $\mu$ with one disorder realization, and an average over the sample complexity as the ensemble average over many disorder realizations. 

For each $\mu$-th training set (or disorder realization), a task vector ${\bf w}^\mu$ is assigned to define the input-output relation in a quasi-linear fashion~\footnote{In principle the relation can be an arbitrary polynomial of $\x_m^\mu$, but here we are confined to a quasi-linear relation for the possibility of analytic treatment.}:

\begin{gather} 
    y^\mu_m = {\bf  w}^\mu \cdot \x_m^\mu  + \sigma \epsilon_m^\mu .
\label{linear-y-vs-x} 
\end{gather}
The input training vectors and the random noise are both Gaussian,
\begin{gather*} 
\x^\mu_m \in {\cal N}(0, I_d /d) \, ,  ~~ 
\epsilon_m^\mu \in {\cal N}(0, 1) \, . 
\end{gather*}
The task vector ${\bf w}^\mu$ is drawn from a pool of task vectors 
\begin{align}
    \Omega = \{ {\bf w}_1 , \cdots, {\bf w}_k \} . 
    \label{w_train}
\end{align}
Each ${\bf w}_p$ ($1 \le p \le k$) is generated from a certain distribution ${\cal P}^{\rm tr}$: ${\bf w}_p \sim {\cal P}^{\rm tr}$. A typical choice is ${\cal P}^{\rm tr} = {\cal N}(0,I_d)$ but others are possible. The number of different task vectors $k$ in $\Omega$ is called the {\it task diversity}. When task diversity $k$ is less than the sample complexity $n$, each task vector may be drawn more than once in the course of training.

For each training set $\mu$, we define the context vector ${\bf H}^\mu$
\begin{align}
{\bf H}^\mu = {\bf x}_{\ell+1}^\mu  \otimes \left( \frac{1}{\ell} \sum_{m=1}^\ell y_m^\mu \widetilde x_m^\mu  \right) ,  ~~ 
\widetilde x_m^\mu \equiv
    \begin{pmatrix}
        d\,\mathbf{x}_m^\mu \\
        y_m^\mu
    \end{pmatrix} \, , 
\label{H-mu}
\end{align}
in accord with the general definition of ${\bf H}$ in \eqref{general-H}. The predicted value for $\hat y^\mu_{\ell+1}$ associated with the $\mu$-th training set is ${\bf P} \cdot {\bf H}^\mu$, which is to be compared with the true output $y^\mu_{\ell+1}$. The difference gives rise to the loss function
\begin{align}
H ( {\bf P} ) = \sum_{\mu=1}^{n}(y_{\ell+1}^\mu - {\bf P}\cdot{\bf H}^\mu)^2 + \frac{n}{d} \lambda\, {\bf P} \cdot {\bf P}
\label{full-batch-loss} 
\end{align}
with a ridge parameter $\lambda$ ($>0$). The gradient-flow dynamics for ${\bf P}$ follows from
\begin{align} 
\frac{\dd\bf P}{\dd t} & \equiv  -\frac{d}{2n} {\bm \nabla}_{\bf P} H({\bf P})  = -{\bm R} {\bf P} + {\bf S} ,  
\label{learning-dynamics} 
\end{align} 
with the {\it relaxation matrix} $\bm R$ and the {\it source vector} $\bf S$ given by
\begin{align} 
{\bm R} & = \lambda I_{d(d+1)} + \frac{d}{n} \sum_{\mu=1}^{n} {\bf H}^\mu ({\bf H}^\mu)^\top , \nn 
{\bf S} & = \frac{d}{n}\sum_{\mu=1}^{n} y_{\ell+1}^\mu {\bf H}^\mu . 
\label{R-and-S} 
\end{align}
The equilibrium ($t\rightarrow \infty$) solution 
\begin{align}
{\bf P}_\infty = {\bm R}^{-1} {\bf S} 
\label{P-infinity} 
\end{align}
gives the learning vector at the end of training. 

Taking the inverse of ${\bm R}$ assumes that all its eigenvalues are non-zero, which is not necessarily guaranteed in the ridgeless limit $\lambda=0$ where
\begin{align} {\bm R} \xrightarrow{\lambda \rightarrow 0^+} {\bm R}_0 = \frac{d}{n} \sum_{\mu=1}^{n} {\bf H}^\mu ({\bf H}^\mu)^\top .   
\label{ridgeless_relaxation}
\end{align} 
The $d(d+1)$-dimensional vectors $\{ {\bf H}^\mu \}_{\mu=1}^n$ are random and almost surely linearly independent if $n\le d(d+1)$. By the standard theory of linear algebra, the non-zero eigenvalues of $(n/d){\bm R}_0$ are the same as those of the $n\times n$-dimensional Gram matrix $\Gamma\in\R^{n\times n}$, whose elements are $\Gamma_{\mu\nu}\equiv{\bf H}^\mu\cdot{\bf H}^\nu$. $\Gamma$ is non-degenerate if and only if ${\bf H}^\mu$ are linearly independent. It follows that ${\rm rank}\,{\bm R}_0={\rm rank}\,\Gamma=\min(n,d(d+1))$, and therefore the fraction of zero eigenvalues of ${\bm R}_0$ is
\begin{align}
    \rho_{\mathrm{flat}} \equiv \frac{\dim\ker{\bm R}_0}{d(d+1)} 
    = \left(1-\frac{n}{d(d+1)}\right)\Theta\left(1-\frac{n}{d(d+1)}\right).
    \label{eq:flat-density}
\end{align}
The eigenvectors in the kernel space of ${\bf R}_0$ define the {\it non-learning}, or the {\it flat} directions because the gradient of the loss function along those directions is zero. The gradient flow preserves any initial component of the learning vector ${\bf P}_\infty$ in the flat directions. The inverse ${\bm R}^{-1}$ is accordingly taken only in the {\it learning} subspace, spanned by eigenvectors with positive eigenvalues. The flat-direction density of the relaxation matrix $\rho_{\mathrm{flat}}$ will eventually be related to the order parameter in the Landau theory of LICL we develop later. 

In the infinite-context limit $\ell/d\to\infty$, each vector ${\bf H}^\mu$ can be replaced by its context average
\begin{align}
    {\bf H}^\mu \xrightarrow{\ell/d \rightarrow \infty} {\bf x}^\mu_{\ell+1} \otimes v^\mu, \quad v^\mu = \begin{pmatrix}
        {\bf w}^\mu \\ |{\bf w}^\mu|^2/d+\sigma^2
    \end{pmatrix} .
\label{Hmu-mean}
\end{align}
The randomness of ${\bf H}^\mu$ is thus determined by that of the two $d$-dimensional vectors ${\bf x}^\mu_{\ell+1}$ and ${\bf w}^\mu$. While the maximum number of independent vectors for ${\bf x}^\mu_{\ell+1}$ is $d$, that of independent ${\bf w}^\mu$ is limited by the size $k$ of the task pool $\Omega$. As $|{\bf w}^\mu|^2/d+\sigma^2$ is nonlinear in ${\bf w}^\mu$, at most $d+1$ $v^\mu$-vectors can be linearly independent even if the corresponding ${\bf w}^\mu$ are not. Consequently, the maximal number of linearly independent ${\bf H}^\mu$ is $d\min (k,d+1)$, which is less than $d(d+1)$ before taking the infinite-context limit. By the same linear-algebra argument as for the finite-context case, one can conclude that ${\rm rank}\,{\bm R}_0=\min(n, d\min(k,d+1)) = \min(n, dk,d(d+1))$, and the total flat-direction density is modified to
\begin{align}
    \rho_{\mathrm{flat}}\xrightarrow{\ell/d \rightarrow \infty} 1-\min\left(\frac{n}{d(d+1)},\frac{k}{d+1},1\right) .
\label{eq:flat-density-infinite-ell}
\end{align}
The distinction between the finite-context flat-direction density in \eqref{eq:flat-density} and the infinite-context expression in \eqref{eq:flat-density-infinite-ell} will play a crucial role in understanding the crossover phenomena in the Landau theory of LICL later.

As in \cite{pehlevan25}, we are interested in the scaling limit of the theory where all the parameters diverge but their ratios are finite~\cite{pehlevan25}: 
\begin{align}
\alpha = \frac{\ell}{d} , ~~ \kappa = \frac{k}{d} , ~~ \tau = \frac{n}{d^2} . 
\label{JAL}
\end{align} 
From now on we refer to these scaled parameters as the normalized context length ($\alpha$), task diversity ($\kappa$), and sample complexity ($\tau$), respectively. All of them play an important role as parameters in the upcoming Landau theory. 

Once the learning vector ${\bf P}_\infty$ is obtained, its ability to predict the output $y_{\ell+1}$ correctly under new circumstances is tested using the {\it  test context}
\begin{align} \{ {\bf x}^{\rm te}_m,y_m^{\rm te} \}_{m=1}^{\ell+1},
\end{align}
where 
\begin{gather} {\bf x}_m^{\rm te} \sim {\cal N}(0,I_d/d),\quad \epsilon_m^{\rm te} \sim {\cal N}(0,1) \nn 
y_m^{\rm te} ={\bf w}^{\rm te} \cdot {\bf x}_m^{\rm te} +\sigma\epsilon_m^{\rm te} . 
\end{gather} 
The crucial difference from the training context is that the task vectors ${\bf w}^{\rm te}$ used in the testing phase are drawn from their own distribution ${\bf w}^{\rm te} \sim {\cal P}^{\rm te}$, which may well differ from ${\cal P}^{\rm tr}$. For each ${\bf w}^{\rm te}$, the correct output $y^{\rm te}_{\ell+1}$ is to be compared with the predicted one, ${\bf P}_\infty \cdot {\bf H}^{\rm te}$, where 
\begin{align} 
{\bf H}^{\rm te} = {\bf x}^{\rm te}_{\ell+1} \otimes \left( \frac{1}{\ell} \sum_{m=1}^\ell y^{\rm te}_m \widetilde{x}^{\rm te}_m \right),  ~~ 
\widetilde x_m^{\rm te} \equiv
    \begin{pmatrix}
        d\,\mathbf{x}_m^{\rm te} \\
        y_m^{\rm te}
    \end{pmatrix} . 
\label{Ht} 
\end{align} 

A quantitative measure of the test error is given by
\begin{align}
    {\cal E} = \mathbb{E} \left[(y^{\rm te}_{\ell+1} - {\bf P}_\infty \cdot {\bf H}^{\rm te})^2\right] ,
\label{test_error}
\end{align}
where $\mathbb{E}[\dots]$ means taking the average over all random variables in the training/testing sets such as ${\bf x}_m^\mu , {\bf x}_m^{\rm te}$, {\it as well as} over the task vectors in both the training/testing stages. The two-tiered averaging is compactly expressed as
\begin{align}
    \mathbb{E}[\dots] = \mathbb{E}_{\rm te} [ \mathbb{E}_{\rm tr} [ \dots ]] , 
\label{E_tr-E_te}
\end{align} 
where $\mathbb{E}_{\rm tr/te} [\dots]$ averages over the random variables as well as the task vectors in the corresponding training or testing stage. The pretraining task pool $\Omega$ remains fixed throughout the averaging process, which is why Eq.~\eqref{E_tr-E_te} is well-defined even when ${\cal P}^{\rm te}$ depends on $\Omega$. An asymptotic expression that is independent of the choice of $\Omega$ is obtained using random matrix theory.

The test error \eqref{test_error} can be written more explicitly as 
\begin{align}
    {\cal E} =& \mathbb{E}_{\rm te}[(y^{\rm te}_{\ell+1})^2] -2 \mathbb{E}_{\rm tr}[ {\bf P}_\infty ]  \cdot \mathbb{E}_{\rm te} [y^{\rm te}_{\ell+1}{\bf H}^{\rm te}] \nn
    &+ {\rm Tr}\left[\mathbb{E}_{\rm tr}\big[{\bf P}_\infty{\bf P}_\infty^\top\big]\mathbb{E}_{\rm te} \big[{\bf H}^{\rm te}({\bf H}^{\rm te})^\top\big]\right].
\label{test_error_ext}
\end{align}
The test average $\mathbb{E}_{\rm te} [\cdots ]$ can be performed straightforwardly. Performing the training average $\mathbb{E}_{\rm tr}$ is much trickier since ${\bf P}_\infty \equiv {\bm R}^{-1} {\bf S}$ involves the inverse of the random matrix $\bm R$. A most straightforward scheme is 
\begin{align} 
\mathbb{E}_{\rm tr}[ {\bf P}_\infty ] & \rightarrow  \bigl( \mathbb{E}_{\rm tr}[ {\bm R} ] \bigr)^{-1} \mathbb{E}_{\rm tr}[ {\bf S} ] , \nn 
\mathbb{E}_{\rm tr} \big[{\bf P}_\infty{\bf P}_\infty^\top\big] & \rightarrow \mathbb{E}_{\rm tr} \big[{\bf P}_\infty ] \mathbb{E}_{\rm tr} \big[ {\bf P}_\infty^\top\big] . 
\end{align} 
We refer to such scheme as {\it annealing}, in analogy with similar use of the terminology in the physics of disorder. A more sophisticated approach will try to  evaluate $\mathbb{E}_{\rm tr} [ {\bf P}_\infty ]$ and $\mathbb{E}_{\rm tr} \big[{\bf P}_\infty{\bf P}_\infty^\top\big]$ as accurately as possible without relying on the above approximation. One such scheme called the {\it cavity method} has been exploited in \cite{pehlevan25}. In that scheme, which we will refer to as {\it quenching}, one finds $\mathbb{E}_{\rm tr} [ {\bf P}_\infty ] \neq \bigl( \mathbb{E}_{\rm tr}[ {\bm R} ] \bigr)^{-1} \mathbb{E}_{\rm tr}[ {\bf S} ]$ and furthermore
\begin{align} 
{\rm Cov} ( {\bf P}_\infty )  \equiv \mathbb{E}_{\rm tr} \big[{\bf P}_\infty{\bf P}_\infty^\top\big] - \mathbb{E}_{\rm tr} \big[{\bf P}_\infty ] \mathbb{E}_{\rm tr} \big[ {\bf P}_\infty^\top\big]  \neq 0 . 
\label{connected-corr-fun-for-P} 
\end{align} 
In the language of statistical physics, this is saying that sample-to-sample fluctuation effects are being captured in the quenching scheme. The covariance of ${\bf P}_\infty$ {\it diverges} at the critical value of the sample complexity $\tau_c = 1$~\cite{pehlevan25}, suggesting the analogy to critical phenomena and the interpretation of sample complexity $\tau$ as the temperature. This observation is what prompted us to construct the Landau theory of LICL.

\section{Annealing scheme} 
\label{sec:annealing}

\subsection{Calculating annealed averages and learning vectors}
The annealing scheme can be viewed as solving the gradient-flow equation \eqref{learning-dynamics} under the approximation ${\bm R} \rightarrow \mathbb{E}_{\rm tr} [{\bm R} ]$, ${\bf S} \rightarrow \mathbb{E}_{\rm tr} [ {\bf S} ]$: 
\begin{align}
    \left( \frac{\dd}{\dd t} + \mathbb{E}_{\rm tr}[{\bm R}] \right) {\bf P}(t)  = \mathbb{E}_{\rm tr}[{\bf S}] . 
\label{annealed-dynamics}
\end{align}
The equilibrium solution is
\begin{align}
    {\bf P}^A_\infty = \Bigl( \mathbb{E}_{\rm tr}[{\bm R}] \Bigr)^{-1}\mathbb{E}_{\rm tr}[{\bf S}] .
\label{P_annealed}
\end{align}

The two averages $\mathbb{E}_{\rm tr}[{\bm R}]$ and $\mathbb{E}_{\rm tr}[{\bf S}]$ are computed in the so-called joint asymptotic limit~\cite{pehlevan25} where all the scaled parameters in \eqref{JAL} are held fixed as we take $d\to\infty$. Since $n$ scales with $d^2$ but $k$ only scales as $d$, the joint asymptotic limit automatically implies $n \gg k$, so that each training task vector ${\bf w}_p$ gets picked $O(d) \gg 1$ times. This allows to replace the empirical average with the statistical one: 
\begin{align}
    \frac{1}{n}\sum_{\mu=1}^n f({\bf w}^\mu ) \rightarrow \frac{1}{k}\sum_{p=1}^k f({\bf w}_p ) . 
\label{n-to-k-main}
\end{align}
For any quantity $f({\bf w})$ that depends on the task vector ${\bf w}^\mu$, one can evaluate the empirical average on the left side with the average $k^{-1} \sum_{p=1}^k$, which can be evaluated analytically for a given distribution ${\cal P}^{\rm tr}$ of the training task vectors. 

When ${\cal P}^{\rm tr} = {\cal N} (0, I_d)$, we can show (see App.~\ref{app:annealed-R-and-S} for derivation) 
\begin{align}
    \mathbb{E}_{\rm tr}[{\bm R}] &= I_d \otimes \big( E^{\rm tr} + \lambda I_{d+1} \big), \nn
    \mathbb{E}_{\rm tr}[{\bf S}] & = \frac{1}{k}\sum_{p=1}^k {\bf w}_p \otimes \begin{pmatrix} {\bf w}_p \\ 1+\sigma^2\end{pmatrix} \nn 
    & = {\rm vec}\left[\begin{pmatrix} {\bm C}^{\rm tr} & (1+\sigma^2){\bf b}^{\rm tr} \end{pmatrix}\right] ,
\label{R-S_ave}
\end{align}
where
\begin{align}
    {\bf b}^{\rm tr}
    \equiv
    \frac{1}{k}\sum_{p=1}^k {\bf w}_p,
    \qquad
    {\bm C}^{\rm tr}
    \equiv
    \frac{1}{k}\sum_{p=1}^k {\bf w}_p{\bf w}_p^\top  , 
    \label{btr-and-Ctr}
\end{align}
are the mean and covariance of the task vectors for a given $\Omega$ in Eq.~\eqref{w_train}, and 
\begin{align}
 E^{\rm tr} =  \begin{pmatrix}
        \gamma I_d+{\bm C}^{\rm tr} & (1+\sigma^2) {\bf b}^{\rm tr} \\
        (1+\sigma^2) ({\bf b}^{\rm tr})^\top & (1+\sigma^2)^2
    \end{pmatrix} ,  \quad
    \gamma =  \frac{1+\sigma^2}{\alpha} . 
\label{E_train}
\end{align}
The matrix vectorization convention ${\rm vec}[{\bf u}v^\top] = {\bf u}\otimes v$
was used in reaching the last line of Eq.~\eqref{R-S_ave}. The annealed optimal learning vector ${\bf P}^A_\infty$ in Eq.~\eqref{P_annealed} becomes
\begin{align}
    {\bf P}^A_\infty =& \left[I_d \otimes \big( E^{\rm tr} + \lambda I_{d+1} \big)^{-1}\right]{\rm vec}\left[\begin{pmatrix} {\bm C}^{\rm tr} & (1+\sigma^2){\bf b}^{\rm tr} \end{pmatrix}\right] \nn
    =& {\rm vec}\left[\begin{pmatrix} {\bm C}^{\rm tr} & (1+\sigma^2){\bf b}^{\rm tr} \end{pmatrix}\big( E^{\rm tr} + \lambda I_{d+1} \big)^{-1}\right] .
\label{PAinfty}
\end{align}
The sample complexity $\tau$ disappears from ${\bf P}^A_\infty$ and all subsequent calculations in the annealing scheme. The $\tau$-dependence is restored in the quenching scheme, reproducing the annealing-scheme result in the $\tau\to\infty$. In the Landau theory, $\tau$ plays the role of temperature and $\tau\to\infty$ corresponds to the infinite-temperature limit where traditionally mean-field-type theories are known to work and critical fluctuations vanish. 

The spectral properties of the covariance matrix ${\bm C}^{\rm tr}$ play a vital role in both annealing and quenching schemes. In the spectral representation,
\begin{align}
    {\bm C}^{\rm tr} = \frac{1}{k}\sum_{p=1}^k{\bf w}_p{\bf w}_p^\top = \sum_{j=1}^d \lambda_j {\bf m}_j {\bf m}_j^\top , 
    \label{Ctr-diag}
\end{align}
where $\{{\bf m}_j\}_{j=1}^d$ is an orthonormal eigenbasis with eigenvalues $\{ \lambda_j \}_{j=1}^d$. ${\bm C}^{\rm tr}$ is a Wishart matrix, whose eigenvalue distribution converges to the Marchenko--Pastur (MP) distribution as $d\to \infty$. Relevant details from random matrix theory are summarized in App.~\ref{app:Wishart-and-MP}. Since ${\rm rank}({\bm C}^{\rm tr}) = \min(k,d)$ almost surely, the fraction of zero eigenvalues of the $d\times d$ matrix $C^{\rm tr}$ is given by $\max(1-\kappa,0)$. The eigenvalues are normalized according to
\begin{align}
    \frac{1}{d}\sum_{j=1}^d\lambda_j = \frac{1}{d}{\rm Tr}\,{\bm C}^{\rm tr} = \frac{1}{dk}\sum_{p=1}^k {\bf w}_p \cdot{\bf w}_p \longrightarrow 1
    \label{normalizing-C-tr}
\end{align}
in the joint asymptotic limit. This normalization convention will be used in both annealing and quenching schemes. 

\subsection{General error estimation} 
The general definition of error introduced in Eq.~\eqref{test_error_ext} reduces, under the annealing scheme, to 
\begin{align}
    {\cal E}^A = & \mathbb{E}_{\rm te}[(y^{\rm te}_{\ell+1})^2] -2{\bf P}^A_\infty \cdot \mathbb{E}_{\rm te} [y^{\rm te}_{\ell+1}{\bf H}^{\rm te}] \nn
    & +({\bf P}^A_\infty)^\top\mathbb{E}_{\rm te} [{\bf H}^{\rm te}({\bf H}^{\rm te})^\top]{\bf P}^A_\infty  . 
\label{annealed_test_error}
\end{align}
The statistical properties of the test context are encoded through the mean and the covariance matrix of the test task distribution ${\cal P}^{\rm te}$:
\begin{align}
    {\bf b}^{\rm te} \equiv \mathbb{E}_{{\bf w}^{\rm te}\sim{\cal P}^{\rm te}}[{\bf w}^{\rm te}], \quad 
    {\bm C}^{\rm te} \equiv \mathbb{E}_{{\bf w}^{\rm te}\sim{\cal P}^{\rm te}}[{\bf w}^{\rm te}({\bf w}^{\rm te})^\top].
\end{align}

Using Eq.~\eqref{PAinfty} for ${\bf P}_\infty^A$, the annealed test error ${\cal E}^A$ can be expressed as
\begin{widetext}
\begin{align}
    {\cal E}^A =& 1+\sigma^2 - \frac{2}{d}{\rm Tr} \left[ \begin{pmatrix} {\bm C}^{\rm tr} & (1+\sigma^2){\bf b}^{\rm tr} \end{pmatrix}
    (E^{\rm tr}+\lambda I_{d+1})^{-1}
    \begin{pmatrix} {\bm C}^{\rm te} & (1+\sigma^2){\bf b}^{\rm te} \end{pmatrix}^{\top} \right] \nn
    &+ \frac{1}{d}{\rm Tr} \left[ \begin{pmatrix} {\bm C}^{\rm tr} & (1+\sigma^2){\bf b}^{\rm tr} \end{pmatrix}
    (E^{\rm tr}+\lambda I_{d+1})^{-1} E^{\rm te} (E^{\rm tr}+\lambda I_{d+1})^{-1}
    \begin{pmatrix} {\bm C}^{\rm tr} & (1+\sigma^2){\bf b}^{\rm tr} \end{pmatrix}^{\top}
    \right] ,
\label{annealed_test_error_general_b}
\end{align} 
\end{widetext}
where we introduced
\begin{align}
    E^{\rm te} =  \begin{pmatrix}
        \gamma I_d+{\bm C}^{\rm te} & (1+\sigma^2) {\bf b}^{\rm te} \\
        (1+\sigma^2) ({\bf b}^{\rm te})^\top & (1+\sigma^2)^2
    \end{pmatrix}
\end{align}
as an analogue to $E^{\rm tr}$. Details of the computation can be found in App.~\ref{app:test-error-under-A}. Explicit calculation shows that, for the two types of test error that will be discussed shortly, terms containing ${\bf b}^{\rm tr}$ and ${\bf b}^{\rm te}$ are $O(1/d)$ compared to terms that do not, and can be ignored in the joint asymptotic limit (see App. \ref{app:btr=0}). The annealed error formula \eqref{annealed_test_error_general_b} then reduces to a function of the covariance matrices ${\bm C}^{\rm tr}$ and ${\bm C}^{\rm te}$: 
\begin{align}
    {\cal E}^A =& \frac{\widetilde\lambda^2}{d}{\rm Tr}\left[\left({\bm C}^{\rm te}-{\bm C}^{\rm tr}\right)\left({\bm C}^{\rm tr}+\widetilde\lambda I_d\right)^{-2}\right] \nn
    &+\sigma^2 + \frac{\widetilde\lambda}{d}{\rm Tr}\left[{\bm C}^{\rm tr}\left({\bm C}^{\rm tr}+\widetilde\lambda I_d\right)^{-1}\right] \nn
    &- \frac{\lambda}{d}{\rm Tr}\left[({\bm C}^{\rm tr})^2\left({\bm C}^{\rm tr}+\widetilde\lambda I_d\right)^{-2}\right] ,
\label{annealed_test_error_general}
\end{align}
where $\tilde \lambda = \lambda + \gamma$.

\subsection{ICL and IDG error estimation} 
Evaluation of the error ${\cal E}^A$ depends on the specific choices of the covariance matrices $C^{\rm tr}$ and $C^{\rm te}$, which follow from their respective distributions ${\cal P}^{\rm tr}$ and ${\cal P}^{\rm te}$. As in \cite{pehlevan25}, we consider two kinds of ${\cal P}^{\rm te}$ leading to the in-distribution generalization (IDG) error and the in-context learning (ICL) error:
\begin{align}
    & \text{IDG: } {\cal P}^{\rm te} = \mathsf{Unif}(\Omega) ~~ \Rightarrow    {\bm C}^{\rm te} = C^{\rm tr}, \nn
    & \text{ICL: } {\cal P}^{\rm te} = \mathcal{N}(0,I_d) ~\, \Rightarrow 
    {\bm C}^{\rm te} = I_d .
\label{IDG-ICL}
\end{align}
Loosely put, the IDG case with ${\bm C}^{\rm te} = {\bm C}^{\rm tr}$ tests how well the pretrained tasks have been memorized, whereas the ICL case with ${\bm C}^{\rm te} \neq {\bm C}^{\rm tr}$ tests the network's performance for unknown tasks. 

For IDG, the first line in the error \eqref{annealed_test_error_general} vanishes:
\begin{align}
    {\cal E}^A_{\rm IDG}
    =& \sigma^2 + \frac{\widetilde\lambda}{d}{\rm Tr}\left[{\bm C}^{\rm tr}\left({\bm C}^{\rm tr}+\widetilde\lambda I_d\right)^{-1}\right] \nn
    &- \frac{\lambda}{d}{\rm Tr}\left[({\bm C}^{\rm tr})^2\left({\bm C}^{\rm tr}+\widetilde\lambda I_d\right)^{-2}\right] .
\end{align}
For the ICL error, 
\begin{align}
    {\cal E}^A_{\rm ICL}
    =& \sigma^2 + \frac{\widetilde\lambda^2}{d}{\rm Tr}\left[\left({\bm C}^{\rm tr}+\widetilde\lambda I_d\right)^{-2}\right] \nn
    &+ \frac{\gamma}{d}{\rm Tr}\left[({\bm C}^{\rm tr})^2\left({\bm C}^{\rm tr}+\widetilde\lambda I_d\right)^{-2}\right].
\label{E_annealed_trace}
\end{align}
Both errors are conveniently expressed using the spectral representation of $C^{\rm tr}$ given in Eq.~\eqref{Ctr-diag}:
\begin{align}
    {\cal E}^A_{\rm IDG} &= \sigma^2 + \frac{\widetilde\lambda}{d}\sum_j \frac{\lambda_j}{\lambda_j+\widetilde\lambda} - \frac{\lambda}{d} \sum_j \bigg(\frac{\lambda_j}{\lambda_j+\widetilde\lambda}\bigg)^2, \nn 
    {\cal E}^A_{\rm ICL} & = \sigma^2 + \frac{1}{d} \sum_j \bigg(\frac{\widetilde\lambda}{\lambda_j+\widetilde\lambda}\bigg)^2 + \frac{\gamma}{d} \sum_j \bigg(\frac{\lambda_j}{\lambda_j + \widetilde\lambda}\bigg)^2 \, . 
\label{E_annealed-spectral}
\end{align}
The sum $\sum_j$ is over both nonzero and zero eigenvalues of $C^{\rm tr}$. 

Various spectral sums in Eq.~\eqref{E_annealed-spectral} can be rewritten using the Stieltjes transform (the resolvent) of the MP distribution~\cite{bai10}, 
\begin{align}
    I_\kappa(z)&=\lim_{\substack{k,d\rightarrow\infty \\ k/d=\kappa}}\frac{1}{d}\sum_{j=1}^d\frac{1}{\lambda_j+z}
    \nn
    &=\frac{2}{z+1-1/\kappa+\sqrt{(z+1-1/\kappa)^2+4z/\kappa}
    } . 
    \label{Stieltjes-main}
\end{align}
%
%
The errors become
\begin{align}
    {\cal E}^A_{\rm IDG} =& \sigma^2 + \widetilde\lambda[1-\widetilde\lambda I_\kappa(\widetilde\lambda)] - \lambda [1 -2 \widetilde\lambda \, I_\kappa(\widetilde\lambda) - \widetilde\lambda^2 I'_\kappa(\widetilde\lambda)] ,\nn 
    {\cal E}^A_{\rm ICL} =& \sigma^2 -1+2\widetilde\lambda I_\kappa(\widetilde\lambda) + (\gamma +1) [1 -2 \widetilde\lambda \, I_\kappa(\widetilde\lambda) - \widetilde\lambda^2 I'_\kappa(\widetilde\lambda)]  . 
\label{E_annealed-Stieltjes}
\end{align}
See App.~\ref{app:derivation_of_annealed_errors} for derivation. These expressions do not depend on the specific choice of the task pool $\Omega$ and only on the task diversity $\kappa$. One can show ${\cal E}^A_{\rm ICL}-{\cal E}^A_{\rm IDG} \ge 0$, which is understandable since IDG draws test task vectors from the same batch as those used in the training. The two annealed error functions in Eq.~\eqref{E_annealed-Stieltjes} do not diverge for any value of the parameters. However, there is a nonanalytic singularity in the ICL error that deserves some discussion. 

In the limit $\widetilde \lambda \to 0^+$, we can show
\begin{align}
    \widetilde\lambda\left({\bm C}^{\rm tr}+\widetilde\lambda I_d\right)^{-1} & =I_d-{\bm C}^{\rm tr}\left({\bm C}^{\rm tr}+\widetilde\lambda I_d\right)^{-1} \nn & \xrightarrow{\tilde \lambda \to 0^+} {\bm \Pi}_0  , 
\label{zero-eigenspace-projector}
\end{align}
where ${\bm \Pi}_0$ is the projector onto the zero-eigenvalue subspace of ${\bm C}^{\rm tr}$ (see App.~\ref{app:derivation_of_annealed_errors} for the proof) with the property $$\frac{1}{d}{\rm Tr}[{\bm \Pi}_0] = \max(1-\kappa,0).$$ This is precisely the fraction of zero eigenvalues of the matrix $C^{\rm tr}$; it vanishes when $C^{\rm tr}$ becomes full-rank. Inserting the projector in Eq.~\eqref{zero-eigenspace-projector} into the annealed error formula \eqref{annealed_test_error_general} and taking the limit $\widetilde\lambda\to0^+$ yields  
\begin{align}
    {\cal E}^A_{\rm IDG} \longrightarrow \sigma^2 ,  \quad
    {\cal E}^A_{\rm ICL} \longrightarrow \sigma^2 + \max(1-\kappa,0) .
\label{annealed-cusp}
\end{align}
The same expressions are obtained by expanding Eq.~\eqref{E_annealed-Stieltjes} to the leading order in $\widetilde \lambda$. While the IDG error shows no dependence on the task diversity $\kappa$ in the limit $\widetilde\lambda=0$, the ICL error develops an additional $\kappa$-dependent, nonanalytic contribution for $0 < \kappa \le 1$. The derivation of Eq.~\eqref{annealed-cusp} via the projector ${\bm \Pi}_0$ demonstrates that the origin of this nonanalyticity is the distribution mismatch [first line on the right-hand side of Eq.~\eqref{annealed_test_error_general}], which vanishes for the IDG error.

\section{Quenching scheme} 
The cavity technique employed in \cite{pehlevan25} is a more sophisticated way to compute the average ${\bf P}_\infty^Q = \mathbb{E}_{\rm tr}  [ {\bm R}^{-1} {\bf S}]$ which differs from the product of averages, $\mathbb{E}_{\rm tr} [ {\bm R} ]^{-1} \mathbb{E}_{\rm tr} [ {\bf S}]$.  Additionally, the average of the product $\mathbb{E}_{\rm tr} \big[{\bf P}_\infty{\bf P}_\infty^\top\big]$ no longer equals the product of averages, ${\bf P}_\infty^Q ( {\bf P}_\infty^Q )^\top$. The latter feature, in particular, leads to the singular behavior in the error at the critical value of sample complexity $\tau_c =1$. Dubbed the {\it double-descent}~\cite{pehlevan25}, the feature bears clear resemblance to the critical phenomena in statistical physics. 

The quenching scheme, i.e. the cavity method, gives the learning vector~\cite{pehlevan25}:
\begin{align}
    {\bf P}^Q_\infty & = \mathbb{E}_{\rm tr} [ {\bm R}^{-1} {\bf S} ] \nn 
    & = {\rm vec}\left[
    \begin{pmatrix}
        {\bm C}^{\rm tr} & (1+\sigma^2){\bf b}^{\rm tr}
    \end{pmatrix}
    (E^{\rm tr}+\xi I_{d+1})^{-1}\right] .
\label{P_infty-quenched}
\end{align}
Compared to the analogous expression in the annealing scheme in \eqref{PAinfty}, the quenched averages ${\bf P}_\infty^Q$ has the renormalized ridge parameter $\xi$ replacing the bare ridge parameter $\lambda$. It is determined self-consistently by the equation~\cite{pehlevan25}
\begin{align}
    \xi^2 I_\kappa(\gamma+\xi)- \tau\lambda= \xi ( 1-\tau ) ,
\label{SC-xi}
\end{align}
with $I_\kappa (z)$ introduced in \eqref{Stieltjes-main}. In the $\tau \rightarrow \infty$ limit, the self-consistency equation gives $\xi=\lambda$ and ${\bf P}_\infty^Q$ reduces to ${\bf P}_\infty^A$. The renormalized ridge parameter $\xi$ is finite for all $\tau$, which ensures the finiteness of ${\bf P}^Q_\infty$ as well.

The test error in Eq.~\eqref{test_error_ext} under the quenching scheme can be expressed as
\begin{align}
    {\cal E}^Q =& {\cal E}_{0}^Q + {\cal E}_1^Q , \nn
    {\cal E}_{0}^Q =& \mathbb{E}_{\rm te}[(y^{\rm te}_{\ell+1})^2] -2{\bf P}_\infty^Q \cdot \mathbb{E}_{\rm te} [y^{\rm te}_{\ell+1}{\bf H}^{\rm te}] \nn
    & +({\bf P}_\infty^Q)^\top\mathbb{E}_{\rm te} [{\bf H}^{\rm te}({\bf H}^{\rm te})^\top]{\bf P}_\infty^Q , \nn
    {\cal E}_1^Q =& {\rm Tr}\left[{\rm Cov}({\bf P}_\infty)\mathbb{E}_{\rm te}\big[{\bf H}^{\rm te} ({\bf H}^{\rm te})^\top\big]\right]
\label{quenched_error-MF-fluct}
\end{align}
where ${\rm Cov}({\bf P}_\infty)$ is the covariance of ${\bf P}_\infty$ given in Eq.~\eqref{connected-corr-fun-for-P}. The test average calculations ${\mathbb E}_{\rm te}[ \cdots ]$ are identical to the annealing case and need not be repeated. Zeroth-order error ${\cal E}^Q_0$ remains finite as ${\bf P}^Q_\infty$ is finite. The fluctuation part ${\cal E}_1^Q$ is given by~\cite{pehlevan25} 
\begin{align}
    {\cal E}_1^Q
    =& \frac{c_e}{d}
    {\rm Tr}\left[E^{\rm te}\left((E^{\rm tr}+\xi I_{d+1})^{-1}
        -\xi(E^{\rm tr}+\xi I_{d+1})^{-2}\right)\right] , \nn[3pt]
    c_e =& \frac{{\cal A}(\xi)}{\tau-1+2\xi I_\kappa(\gamma+\xi)+\xi^2 I'_\kappa(\gamma+\xi)} , \nn[3pt]
    \mathcal A(\xi) =& \sigma^2+\gamma+\xi-(\gamma+\xi)^2I_\kappa(\gamma+\xi) \nn
    &-\xi\left[1-2(\gamma+\xi)I_\kappa(\gamma+\xi) -(\gamma+\xi)^2I_\kappa'(\gamma+\xi)\right].
    \label{quenched_error-fluct}
\end{align}
The overall degree of fluctuation is captured by the scalar $c_e$. In the limit $\tau\to\infty$, the fluctuation contribution ${\cal E}_1^Q$ vanishes, while the zeroth-order contribution ${\cal E}_0^Q$ reduces to the annealed error ${\cal E}^A$. This is the limit where $n = d^2 \tau$, the number of pretraining sample contexts, far exceeds the number of trainable parameters in ${\bf P}$. In this case, the empirical averages over $n$ samples in the definitions of the relaxation matrix ${\bm R}$ and the source vector ${\bf S}$ [Eq.~\eqref{R-and-S}] converge to their distribution averages $\mathbb{E}_{\rm tr}[\dots]$, so that the full quenched learning dynamics of Eq.~\eqref{learning-dynamics} is essentially captured by the annealed dynamics of Eq.~\eqref{annealed-dynamics}. In the Landau-theory interpretation, $\tau \to \infty$ corresponds to the infinite-temperature limit where connected correlations vanish.

Substituting the test task distributions in Eq.~\eqref{IDG-ICL} into Eq.~\eqref{quenched_error-fluct} and combining this with the results of the annealing scheme in Eq.~\eqref{E_annealed-Stieltjes} while replacing $\lambda$ by $\xi$, the IDG and ICL errors under quenching are obtained as~\cite{pehlevan25}
\begin{align}
    {\cal E}_{\rm IDG}^Q
    =& \sigma^2+\zeta-\zeta^2I_\kappa(\zeta) \nn
    &- \xi \left[ 1 - 2\zeta I_\kappa(\zeta) - \zeta^2I_\kappa'(\zeta) \right] \nn
    &+ c_e \left[ 1 - 2\xi I_\kappa(\zeta) - \xi^2I_\kappa'(\zeta) \right] , \nn
    {\cal E}_{\rm ICL}^Q
    =& 1+\sigma^2 - 2\left[1-\zeta I_\kappa(\zeta)\right] \nn
    & + (1+\gamma)
    \left[1 - 2\zeta I_\kappa(\zeta) - \zeta^2 I_\kappa'(\zeta) \right] \nn
    &+ c_e(1+\gamma) \left[ I_\kappa(\zeta)+\xi I_\kappa'(\zeta) \right] ,
\label{E_ICL-IDG_quenched}
\end{align}
where $\zeta=\gamma+\xi$. Terms that contain ${\bf b}^{\rm tr}$ and ${\bf b}^{\rm te}$ become negligible in the joint asymptotic limit and do not contribute to Eq.~\eqref{E_ICL-IDG_quenched}.

Taking the limits $\alpha,\tau\to\infty$ simultaneously while fixing the ratio $c^*=\alpha/\tau$ and the ridgeless limit $\lambda=\lim_{\tau\to\infty}\xi=0$, Eq.~\eqref{E_ICL-IDG_quenched} becomes~\cite{pehlevan25}
\begin{align}
    {\cal E}^Q_{\rm IDG} \longrightarrow& \sigma^2 ,  \nn
    {\cal E}^Q_{\rm ICL} \longrightarrow& \sigma^2 + \left(1+\dfrac{\sigma^2}{1+\sigma^2}c^*\right)\max(1-\kappa,0) .
\end{align}
Setting $c^*=0$ reproduces the $\widetilde\lambda=\lambda+\gamma=0$ limit of the annealed errors in Eq.~\eqref{annealed-cusp}, as $c^*=0$ is equivalent to first taking the limit $\tau\to\infty$ and then $\gamma=(1+\sigma^2)/\alpha\to0^+$, and for $\tau\to\infty$, the quenched errors in Eq.~\eqref{E_ICL-IDG_quenched} reduce to their annealed counterparts in Eq.~\eqref{E_annealed-Stieltjes}.

\section{Landau theory of LICL}
The fluctuation part of the error given in Eq.~\eqref{quenched_error-fluct} was obtained for general choice of distributions for the testing task vectors. The divergence of the error encapsulated in the constant $c_e$ is therefore a universal feature of the LICL that calls for a general interpretation. We construct a Landau theory that provides such a framework.  

The self-consistency equation for $\xi$, 
\begin{align} 
\xi (\tau-  1)  + \xi^2 I_\kappa(\gamma+\xi)-\tau\lambda =0 , 
\label{SC-xi-again} 
\end{align} 
follows as the stationary solution $V'(\xi_*) = 0$ of the potential 
\begin{align}
     V =\frac{\tau-1}{2}\xi^2+
     \int_0^\xi {\rm d}u\, u^2 I_\kappa(\gamma+u) - h \xi .
    \label{eq:Landau-potential}
\end{align}
We argue that this is the desired Landau functional for the order parameter $\xi$. 
The sample complexity $\tau$ is the {\it temperature}, while the ridge parameter $\lambda$ plays the role of {\it conjugate field} $h \equiv \tau \lambda$. At zero field $h=0$, the order parameter vanishes at the critical temperature $\tau_c = 1$. Critical exponents can be computed readily from the Landau theory. 

\subsection{Landau theory for finite $\gamma$}
A critical region is where, by definition, the order parameter $\xi$ becomes very small. For finite inverse context length $\gamma$, this critical regime obeys $\gamma \gg \xi$ 
and the Landau functional can be expanded as 
\begin{align}
    V \simeq \frac{t}{2}\xi^2 + \frac{I_\kappa(\gamma)}{3}\xi^3-h\xi , \quad (t\equiv \tau-1) . 
    \label{Landau-expansion-for-large-gamma} 
\end{align}
The cubic term in $\xi$ is allowed because $\xi$, being a non-negative variable by definition, lacks the $\xi \rightarrow -\xi$ symmetry. As a consequence, the order parameter $\xi$ does not bear the ordinary interpretation of spontaneously broken symmetry. We give an alternative, geometric interpretation of the order parameter shortly. The saddle-point equation \eqref{Landau-expansion-for-large-gamma} is solved by
\begin{align}
\xi_*  \simeq \frac{-t +\sqrt{t^2+4 h I_\kappa (\gamma) }}{2 I_\kappa (\gamma) } . 
\end{align}
Two limiting solutions 
\begin{align}
    \xi_*&\simeq\frac{-t}{I_\kappa (\gamma)} \Theta (-t) 
    &(h=0), \nn
    \xi_*&\simeq\left[\frac{h}{I_\kappa (\gamma)}\right]^{1/2}
    &(\tau=1) , 
    \label{general-chi-expressions} 
\end{align}
yield the critical exponents:
\begin{align} 
( \beta_{\rm cr} , \delta_{\rm cr} ) = (1, 2). 
\end{align}

Differentiating the saddle-point equation 
\begin{align*}
V'(\xi_*)=(\tau-1) \xi_*+\xi_*^2 I_\kappa(\gamma+\xi_*)-h=0    
\end{align*}
 with respect to $h$ results in the {\it susceptibility} $\chi$ 
\begin{align}
    \chi \equiv \frac{\partial\xi_*}{\partial h }\bigg|_{\tau, \gamma}  =\frac{1}{V''(\xi_*)},
\label{eq:susceptibility}
\end{align}
where $V''(\xi) =\tau-1 +2\xi I_\kappa(\gamma+\xi) +\xi^2 I_\kappa'(\gamma+\xi)$. Explicit calculation near $\tau = \tau_c$ gives
\begin{equation}
    \chi
    =\frac{1}{\sqrt{t^2+4I_\kappa(\gamma)h}}  ,
    \label{eq:chi-scaling}
\end{equation}
which diverges as $\sim |t|^{-1}$ at $h =0$, resulting in the critical exponent 
\begin{align} \gamma_{\rm cr} =1.
\end{align} 
The three critical exponents $(\beta_{\rm cr}, \delta_{\rm cr}, \gamma_{\mathrm{cr}} ) = (1,2,1)$, obey the Widom scaling relation
\begin{align}
    \gamma_{\mathrm{cr}}=\beta_{\rm cr} (\delta_{\rm cr} -1).
\end{align}

The fluctuation coefficient $c_e$ in \eqref{quenched_error-fluct} is proportional to the susceptibility:
\begin{align}
    c_e=\frac{{\cal A}(\xi_*)}{V''(\xi_*)} = {\cal A}(\xi_*) \chi .
    \label{eq:c_e-chi-relation}
\end{align}
For fixed $\gamma>0$, the critical saddle satisfies $\xi_*\to0$, and
\begin{align}
{\cal A}(\xi_*)\to {\cal A}(0)=\sigma^2+\gamma-\gamma^2 I_\kappa(\gamma)>0.    
\end{align}
Thus, ${\cal A}(\xi_*)$ remains finite and nonzero in the fixed-$\gamma$ critical limit, and $c_e\sim{\cal A}(0)\chi$ inherits the divergence of $\chi$. This conclusion is not uniform in the strict limit $\gamma\to0$.
The phase diagram for finite $\gamma$ is a very simple one, with the critical temperature $\tau_c = 1$ and the associated critical exponents $(\beta_{\rm cr}, \delta_{\rm cr}, \gamma_{\rm cr} ) = (1,2,1)$.

What is the meaning of the order parameter $\xi_*$? From the cavity method~\cite{pehlevan25}, we know that the inverse of the relaxation matrix ${\bm R}$ is asymptotically given by
\begin{align}
    {\bm R}^{-1} = I_d \otimes \frac{\xi}{\lambda}\left(E^{\rm tr}+\xi I_{d+1}\right)^{-1} .
\end{align}
Taking the trace on both sides, we obtain
\begin{align}
    \frac{\lambda}{d^2}{\rm Tr}[{\bm R}^{-1}] =& \frac{\xi}{d}{\rm Tr}\left[(E^{\rm tr}+\xi I_{d+1})^{-1}\right] \nn
    =& \xi I_\kappa(\gamma+\xi) .
\end{align}
Since ${\bm R}={\bm R}_0+\lambda I$, the left-hand side selects in the ridgeless limit $\lambda\to0$ the zero eigenvalues of the ridgeless relaxation matrix ${\bm R}_0$, which gives nothing but its flat-direction density $\rho_{\mathrm{flat}}$ defined in Eq.~\eqref{eq:flat-density}. We have thus established a relation between the order parameter $\xi_*$ and the geometric quantity $\rho_{\mathrm{flat}}$ in the zero-field limit $h=\tau\lambda=0$,
\begin{align}
    \rho_{\mathrm{flat}} = I_\kappa(\gamma+\xi_*)\xi_* .
\label{eq:rho-xi-relation}
\end{align}
Recall that the flat-direction density of ${\bm R}_0$ previously obtained in Eq.~\eqref{eq:flat-density} becomes 
\begin{align}
    \rho_{\mathrm{flat}} = (-t)\Theta(-t)
\label{eq:flat-direction-asymp}
\end{align}
in the joint asymptotic limit. Since $I_\kappa(\gamma+\xi_*)\simeq I_\kappa(\gamma)$ near criticality, plugging Eq.~\eqref{eq:flat-direction-asymp} into the relation \eqref{eq:rho-xi-relation} indeed reproduces the zero-field critical solution in Eq.~\eqref{general-chi-expressions}.

\subsection{Crossover behavior for $\gamma \ll \xi_* \ll 1$}
In the large-context-length regime \(\gamma \ll 1\), there are situations where the order parameter $\xi_*$, though small, is still much larger than $\gamma$. This regime, $\gamma \ll \xi_* \ll 1$, deserves a special attention as it displays significant depression of the order parameter somewhat reminiscent of the crossover behavior. In this {\it crossover regime}, $\gamma + \xi$ is small and the resolvent $I_\kappa (\gamma + \xi)$ can be replaced by its asymptotic expression: 

\begin{equation}
    I_\kappa(z)=
    \begin{cases}
       \dfrac{1-\kappa}{z}+\dfrac{\kappa^2}{1-\kappa}+O(z), & \kappa<1,\\[3mm]
       \dfrac{1}{\sqrt{z}}-\dfrac{1}{2}+O(\sqrt{z}), & \kappa=1,\\[3mm]
       \dfrac{\kappa}{\kappa-1}+O(z), & \kappa>1 . 
    \end{cases}
    \label{eq:mp-asymptotics}
\end{equation}
In the high-task-diversity ($\kappa > 1$) regime, the resolvent $I_\kappa (z) \rightarrow \kappa/(\kappa-1)$ stays finite and all the discussion of critical exponents including the critical temperature $\tau_c =1$ from the previous subsection remains unchanged.  

The low-task-diversity regime $\kappa<1$ is more interesting because the singular behavior of the resolvent leads to the saddle-point equation 
\begin{align}
    (\tau-\kappa)\xi_*-(1-\kappa)\frac{\gamma\xi_*}{\gamma+\xi_*}+\frac{\kappa^2}{1-\kappa}\xi_*^2-h =0 .
\label{eq:low-kappa-saddle}
\end{align}
In the crossover regime $\gamma\ll\xi_*$, solving this equation at $h=0$ gives
\begin{align}
    \xi_* & \simeq \frac{1-\kappa}{2\kappa^2}
    \left[-(\tau-\kappa)
    +\sqrt{(\tau-\kappa)^2+4\kappa^2\gamma}\right] \nn 
    & \xrightarrow{\kappa-\tau \gg \sqrt{\gamma}} \frac{1-\kappa}{\kappa^2}(\kappa-\tau)\Theta(\kappa-\tau) .
    \label{eq:low-kappa-crossover-scaling}
\end{align}
This means that when the temperature $\tau$ is sufficiently below $\kappa$, the critical temperature appears to have shifted to $\kappa < 1$. Indeed, if we take the $\gamma \to 0^+$ limit first, this will be the new transition temperature predicted by the Landau theory. 

\begin{figure}
    \centering

    \makebox[\linewidth][c]{%
        \hspace{-1.48cm}
        \includegraphics[width=0.8\linewidth]{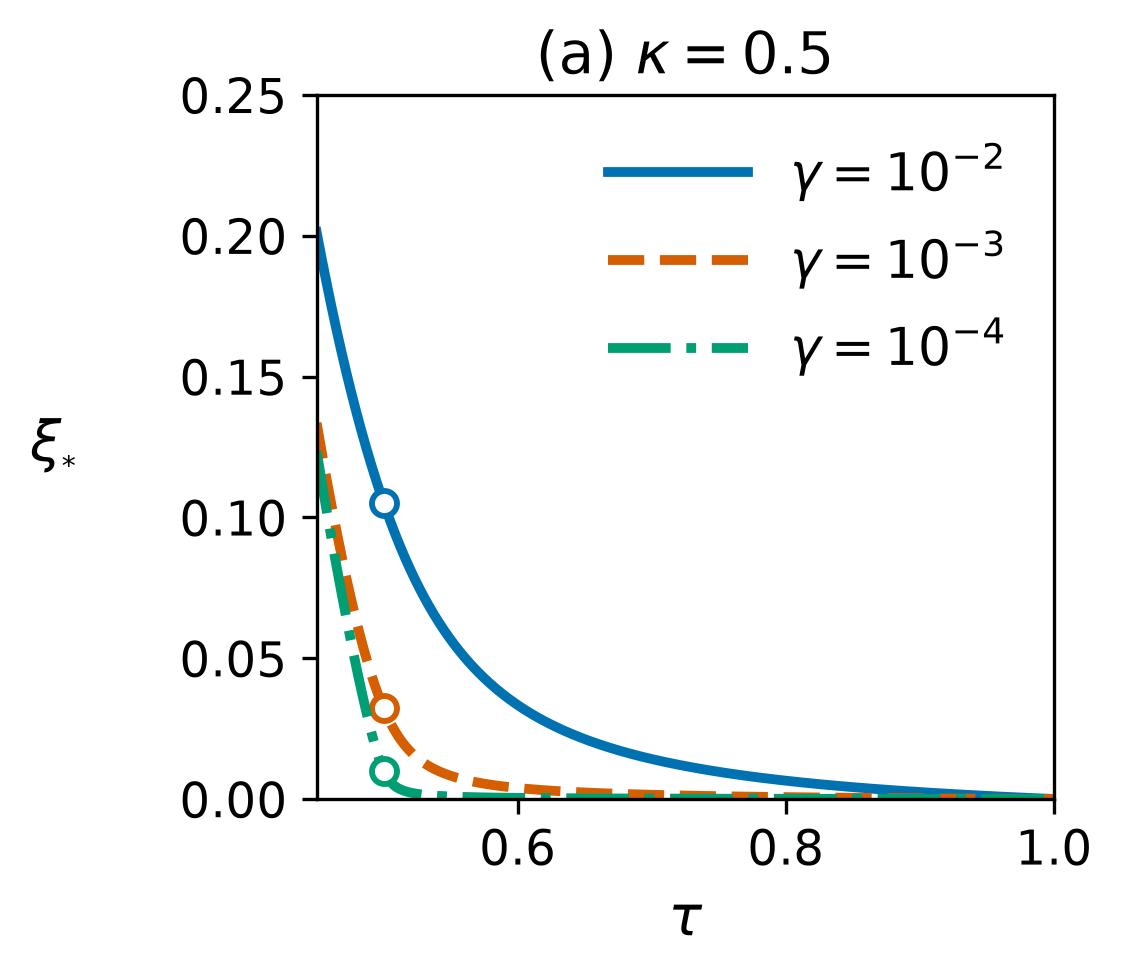}
    }

    \makebox[\linewidth][c]{%
        \hspace{-1.3cm}
        \includegraphics[width=0.8\linewidth]{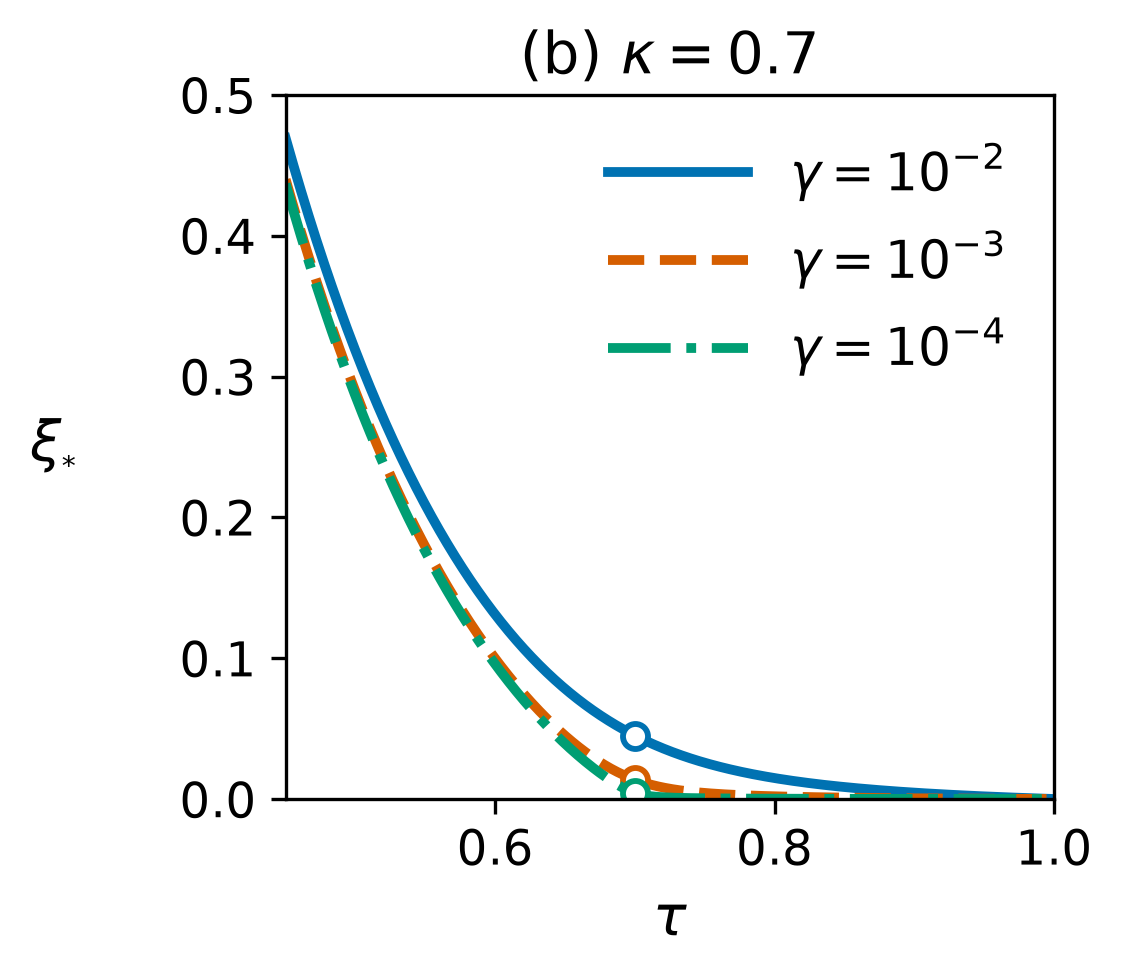}
    }

    \caption{Order parameter $\xi_*$ as a function of $\tau$ below the critical temperature $\tau_c =1$ for several values of $\kappa<1$ and $\gamma\ll1$. The crossover temperature $\tau=\kappa$ is indicated by hollow circles. The residual order in the crossover region $\kappa<\tau<1$ diminishes as $\gamma\to0$.}
    \label{fig:low-kappa}
\end{figure}

Assuming that $\gamma$ is small but still finite, there will eventually be a critical regime $\xi_* \ll \gamma \ll 1$, where the correct saddle-point equation to deal with is
\begin{align}
\xi_* \left[ \tau-1 + \frac{1-\kappa}{\gamma}\xi_* \right] =0 , 
\label{eq:low-kappa-critical}
\end{align}
with the solution
\begin{align}
    \xi_* \simeq \gamma\frac{1-\tau}{1-\kappa}\Theta(1-\tau) . 
\label{eq:low-kappa-true-crit}
\end{align}
The true critical temperature is $\tau_c = 1$. Moreover, one can show that the residual order in the entire \textit{pseudgap region} $\kappa<\tau<1$ has the scaling behavior $\xi_*=O(\gamma)$. Analyzing the saddle-point equation \eqref{eq:low-kappa-saddle} for finite $h$, we find the same critical exponents $(\beta_{\rm cr} , \delta_{\rm cr} , \gamma_{\rm cr}) = (1,2,1)$ in both crossover and critical regimes. Plots of the exact solution to the self-consistent equation \eqref{SC-xi-again} shown in Fig. \ref{fig:low-kappa} illustrate the behavior of $\xi_*$ described above near the crossover and critical temperatures $\tau=\kappa$ and $\tau_c=1$, respectively, as well as how the size of the residual order shrinks as $\gamma\to0$.

The crossover phenomenon can be understood geometrically using the relation \eqref{eq:rho-xi-relation} of the order parameter with the flat-direction density of the relaxation matrix $\rho_{\rm flat}$. In the crossover regime $\gamma\ll\xi_*\ll1$,  Eq.~\eqref{eq:rho-xi-relation} becomes
\begin{align}
    \rho_{\mathrm{flat}} \simeq 1-\kappa + \frac{\kappa^2}{1-\kappa}\xi_* . 
\label{eq:rho-xi-crossover}
\end{align}
On the other hand, the infinite-context limit ($\gamma=0$) of the flat-direction density $\rho_{\mathrm{flat}}$ obtained earlier as Eq.~\eqref{eq:flat-density-infinite-ell} becomes, in the joint asymptotic limit,  
\begin{align}
    \rho_{\mathrm{flat}} = 1-\min\left(\tau,\kappa,1\right) .
\label{eq:flat-density-gamma-0}
\end{align}
For $\kappa < 1$, this reduces to $\rho_{\mathrm{flat}} = 1-\min\left(\tau,\kappa \right)$. Substituting this expression into the left-hand side of Eq.~\eqref{eq:rho-xi-crossover} and rearranging terms, we recover the crossover solution for $\xi_*$ previously obtained in Eq.~\eqref{eq:low-kappa-crossover-scaling}. 

In the critical regime $\xi_*\ll\gamma\ll1$, the relation \eqref{eq:rho-xi-relation} becomes
\begin{align}
    \rho_{\mathrm{flat}} \simeq \frac{1-\kappa}{\gamma}\xi_* .
\label{eq:rho-xi-true-crit}
\end{align}
In this regime, we resort to the finite-context behavior of $\rho_{\mathrm{flat}}$ in Eq.~\eqref{eq:flat-direction-asymp} and thereby reobtain the critical solution \eqref{eq:low-kappa-true-crit} from Eq.~\eqref{eq:rho-xi-true-crit}. Such recovery of expressions for $\xi_*$ bolsters the claim that the origin of the order parameter is indeed geometric. 

The temperature $\tau = \kappa$ defines a true critical temperature only when we take the $\gamma \to 0^+$ limit first. In that regard, the small residual order parameter of size $O(\gamma)$ can be viewed as a result of the fluctuations created by small but finite $\gamma$.

\begin{figure}[]
    \centering
    \hspace{-1.2cm}
    \includegraphics[width=0.8\columnwidth]{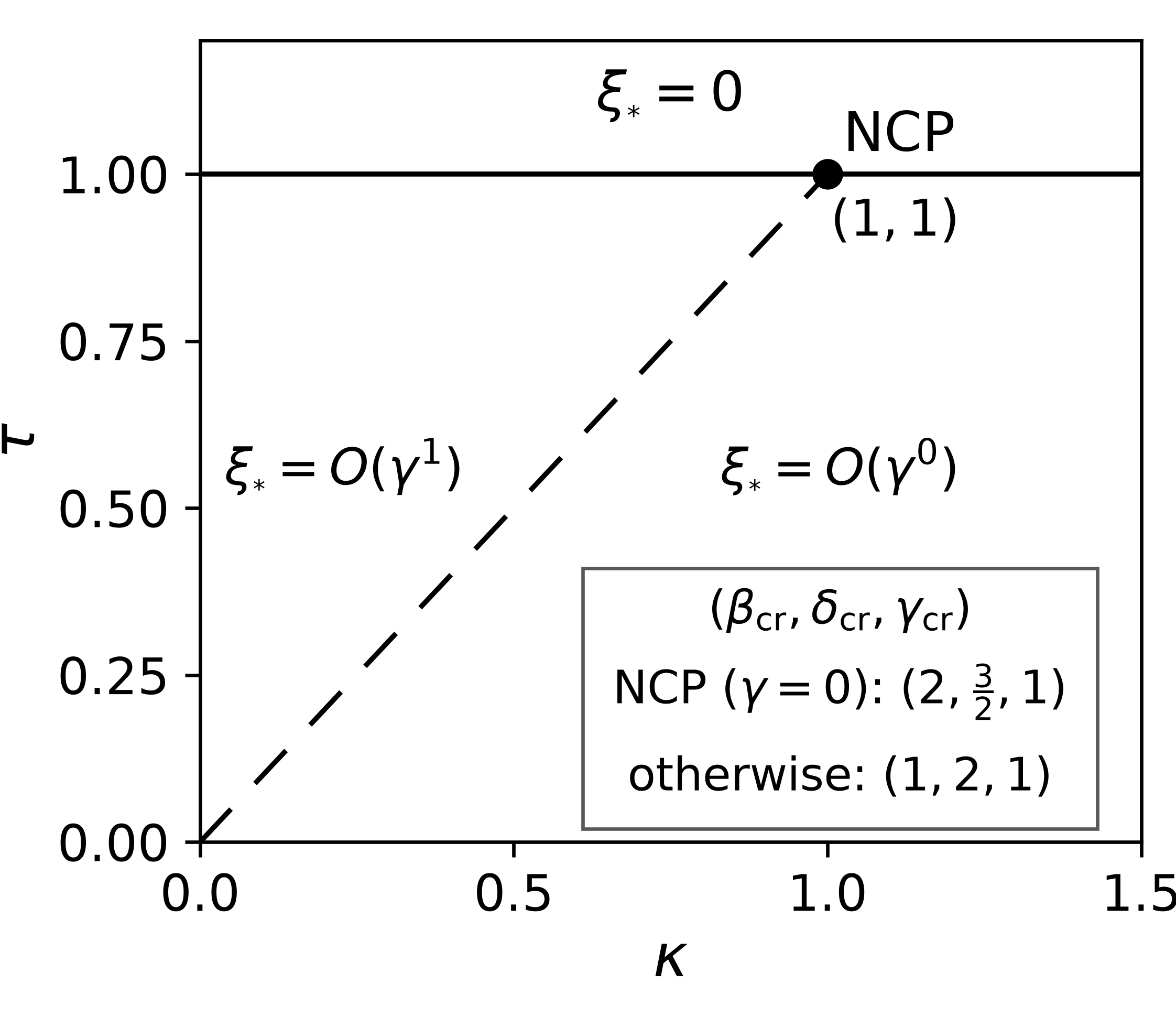}
    \caption{Schematic phase diagram for $\gamma\ll1$ in the $(\kappa,\tau)$ plane. The transition between the ordered phase $\xi_*>0$ and disordered phase $\xi_*=0$ occurs at $\tau_c=1$ (solid line). Above the crossover temperature $ \tau = \kappa$ (dashed line), the ordered phase exhibits a small residual order $\xi_*=O(\gamma)$ characteristic of a pseudogap regime. When we take the $\gamma \to 0$ limit first, the crossover line becomes the exact phase transition. Furthermore, the nonanalytic critical point (NCP) at $(\kappa, \tau)=(1,1)$ develops new critical exponents $(\beta_{\rm cr}, \delta_{\rm cr}, \gamma_{\rm cr}) = (2,3/2,1)$ different from the rest of the phase diagram where $(\beta_{\rm cr}, \delta_{\rm cr}, \gamma_{\rm cr}) = (1,2,1)$. }
    \label{fig:phase-diagram}
\end{figure}

At $\kappa=1$, which marks the endpoint of the pseudogap regime, we use the asymptotic expansion for $I_1 (z)$ in \eqref{eq:mp-asymptotics} to write down the appropriate saddle-point equation, 
\begin{align}
t \xi+\frac{\xi^2}{\sqrt{\gamma+\xi}} -\frac{1}{2}\xi^2-h = 0.
    \label{eq:kappa-one-finite-gamma-saddle}
\end{align}
Since it is assumed that $\gamma + \xi \ll 1$, we can ignore the $-\xi^2/2$ term and solve the equation for $h=0$:
\begin{equation}
\xi_*=\frac{t^2-t\sqrt{t^2+4\gamma}}{2} \simeq
    \begin{cases}
        (-t)\sqrt{\gamma}
        & ( -t\ll\sqrt{\gamma} )
        \\[1mm]
        t^2
        & ( -t\gg\sqrt{\gamma} )
    \end{cases}. 
    \label{eq:kappa-one-crossover}
\end{equation}

As long as $\gamma$ is small but nonzero, the region $-t \ll \sqrt{\gamma}$ corresponds to the critical regime, $\xi_* \ll \gamma$, and the critical exponents are the same as before. For $\gamma=0$, however, the region $-t \ll \sqrt{\gamma}$ vanishes, 
and with the new critical solution $\xi_*\simeq t^2$, the critical exponent $\beta_{\rm cr}$ changes to 2. To obtain other critical exponents in the strict infinite-context limit $\gamma=0$, it is convenient to first derive the Landau functional valid in that regime:
\begin{align}
    V
    \simeq\frac{t}{2}\xi^2+\frac{2}{5}\xi^{5/2}-h\xi .
    \label{eq:multicritical-potential}
\end{align}
Due to the nonanalytic $\xi^{5/2}$ term in the Landau expansion, the critical exponents are now
\begin{align}
    (\beta_{\rm cr},\delta_{\rm cr},\gamma_{\rm cr})
    =\left(2,\frac{3}{2},1\right) . 
\end{align}
They also satisfy the Widom scaling. $(\kappa, \tau ) = (1,1)$ becomes a nonanalytic critical point (NCP) with these critical exponents. A schematic phase diagram valid for $\gamma \ll 1$ is shown in Fig.~\ref{fig:phase-diagram}.

\section{Numerical tests of Landau theory}
The analytic Landau theory predicts two distinct regimes for the order parameter $\xi_*$ in the ordered phase $\tau<1$: an $O(\gamma^0)$ regime and an $O(\gamma^1)$ pseudogap regime. We perform an independent numerical check on this. At each point $(\kappa, \tau)$, we choose $k$ and $n$ as the nearest integers to $\kappa d$ and $\tau d^2$,  respectively, in accordance with the definition in Eq.~\eqref{JAL}. For numerical simulation we choose $d=30$. A given choice of $\gamma$ fixes $\alpha=(1+\sigma^2)/\gamma$ through Eq.~\eqref{E_train}, and then $\ell$ is chosen as the nearest integer to $\alpha d$. 

For each $(\kappa, \tau)$, we generate $B$ independent task pools $\Omega^{(1)},\dots,\Omega^{(B)}$ of the form given in Eq.~\eqref{w_train}, with $B$ up to 1024. Each task vector ${\bf w}_p$ is drawn independently and uniformly from a sphere $S^{d-1} (\sqrt{d})$. For each task pool $\Omega^{(b)}$, we generate $G$ independent realizations of the pretraining data $\{{\bf x}^\mu_1,y^\mu_1,\dots,{\bf x}^\mu_{\ell+1},y^\mu_{\ell+1}\}_{\mu=1}^n$ with the relation of each pair $({\bf x}^\mu_m,y^\mu_m)$ as defined in Eq.~\eqref{linear-y-vs-x}. Each task vector ${\bf w}_p$ is used approximately $[n/k]$ times in the generation process. Each context vector ${\bf H}^\mu$ is generated according to the formula \eqref{H-mu}. For each realization $1 \le b \le B$ and $1 \le g \le G$, the relaxation matrix ${\bm R}^{(b,g)}$ and the source vector ${\bf S}^{(b,g)}$ are constructed according to Eq.~\eqref{R-and-S}, and the stationary learning vector is obtained by solving the linear equation ${\bm R}^{(b, g)}{\bf P}^{(b,g)}_{\infty}={\bf S}^{(b,g)}$. We set $\sigma^2 = 0$ in the numerical simulation.

The quantities ${\bf b}^{{\rm tr}, (b)}$, ${\bm C}^{{\rm tr},(b)}$, and $E^{{\rm tr}, (b)}$ are constructed from a given pool $\Omega^{(b)}$ for each $1 \le b \le B$ using Eqs.~\eqref{btr-and-Ctr} and \eqref{E_train}. A non-negative value of the order parameter is then obtained by 
\begin{align}
\xi_* & =\underset{0\leq\xi}{\operatorname{arg\,min}}\, \frac{1}{B}\sum_{b=1}^B \Bigg\|\frac{1}{G}\sum_{r=1}^G {\bf P}_{\infty}^{(b,g)} \nn
    & -{\rm vec}\left[\begin{pmatrix}
        {\bm C}^{{\rm tr},(b)} & (1+\sigma^2){\bf b}^{{\rm tr},(b)}
    \end{pmatrix}(E^{{\rm tr},(b)}+\xi I_{d+1})^{-1}\right]\Bigg\|_2^2 . 
    \label{eq:numerical-xi-fit}
\end{align}
The formula \eqref{eq:numerical-xi-fit} provides an independent means of extracting $\xi_*$ without relying on the self-consistency equation \eqref{SC-xi}. 
\begin{figure}[]
    \centering

    \makebox[\columnwidth][c]{%
        \hspace{-0.4cm}
        \includegraphics[width=0.8\columnwidth]{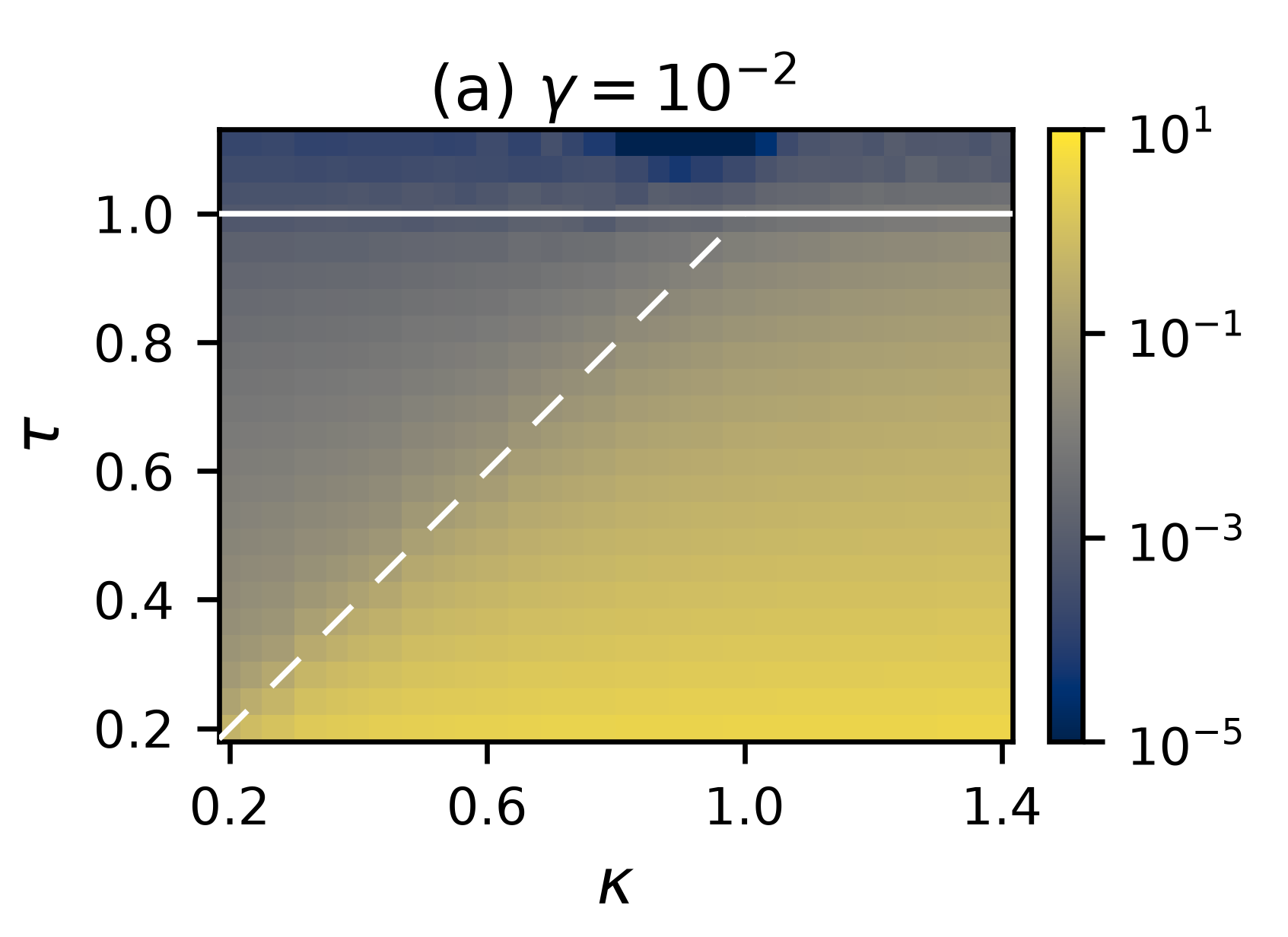}
    }

    \makebox[\columnwidth][c]{%
        \hspace{-0.4cm}
        \includegraphics[width=0.8\columnwidth]{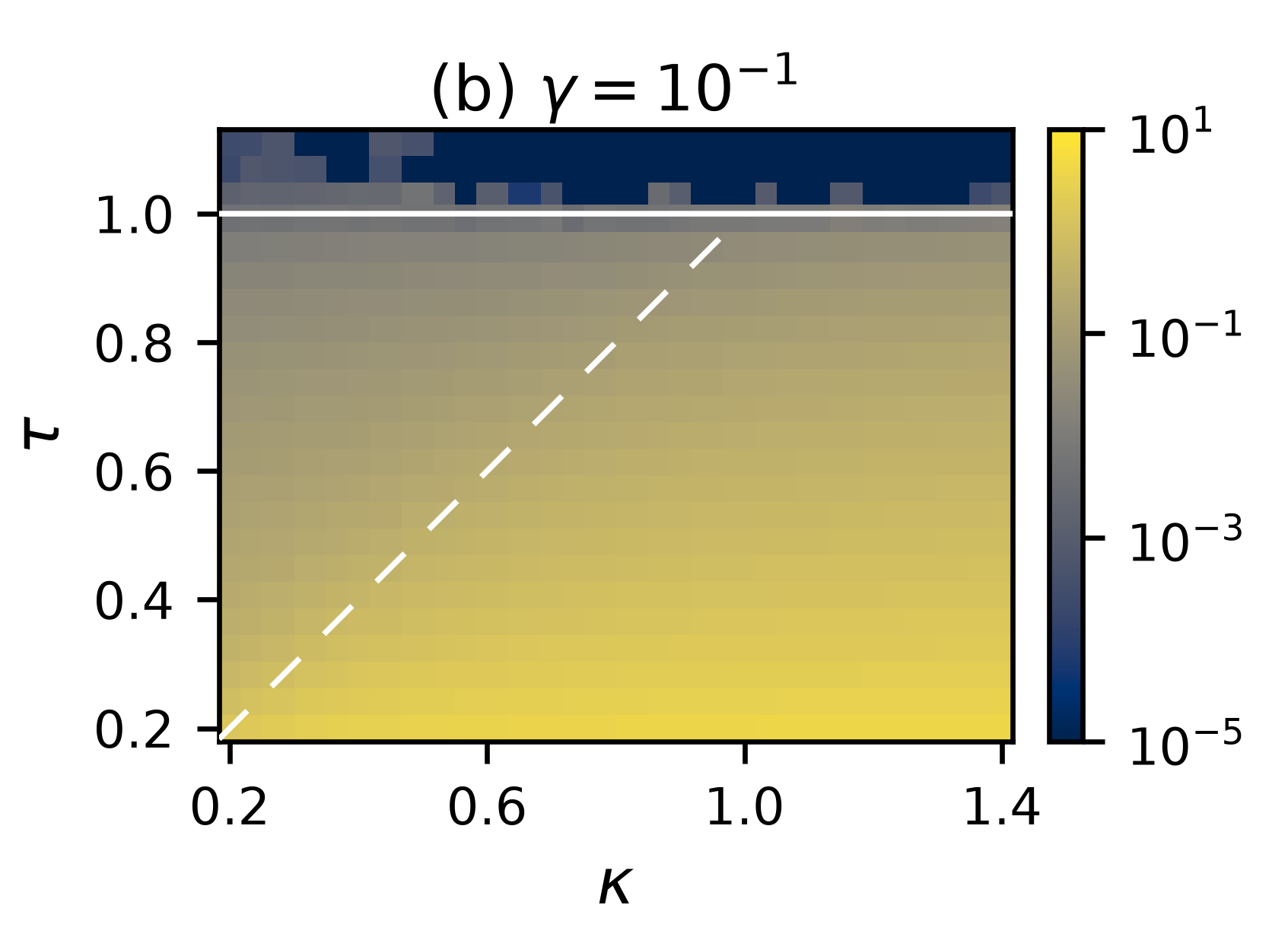}
    }

    \makebox[\columnwidth][c]{%
        \hspace{-1.9cm}
        \includegraphics[width=0.78\columnwidth]{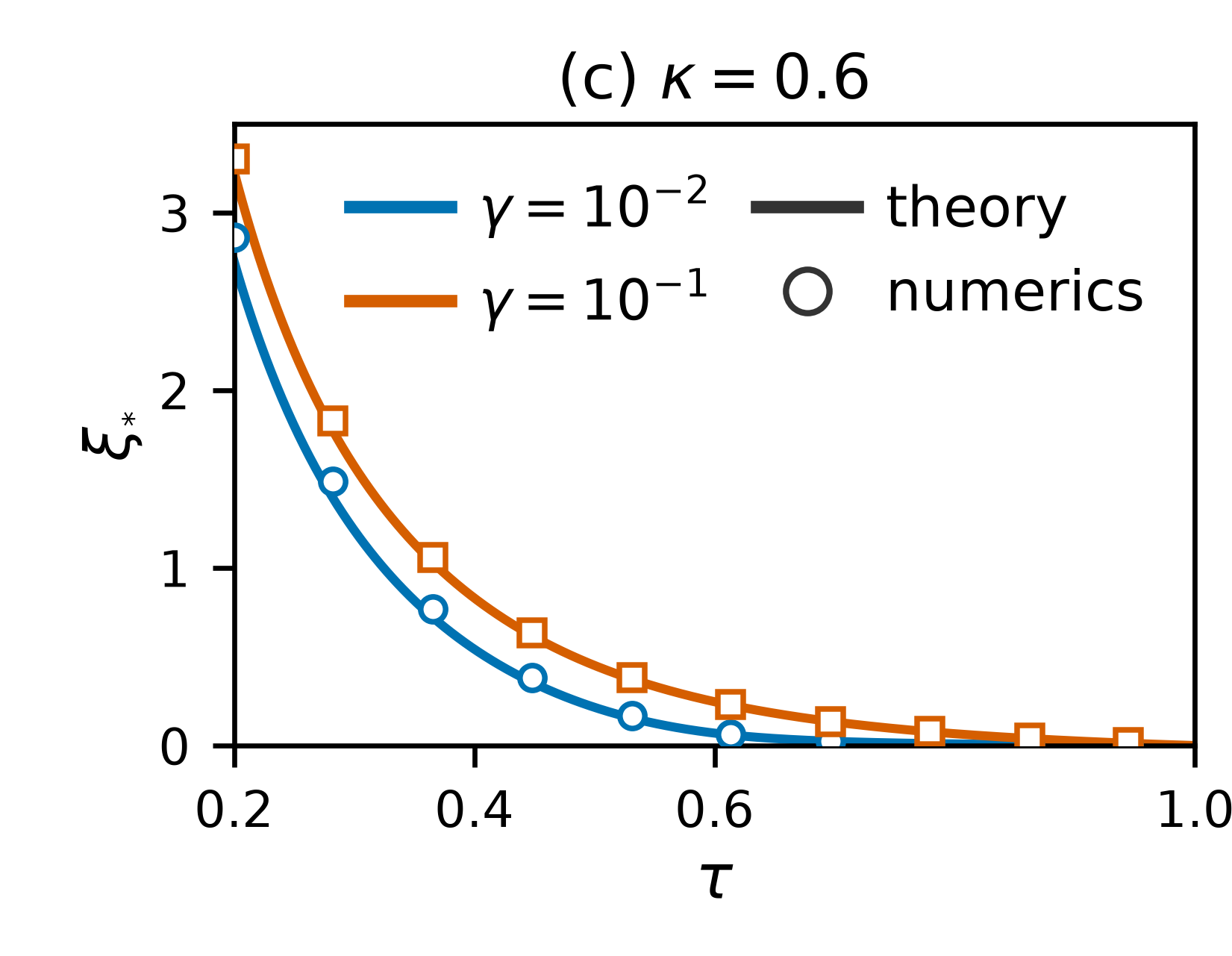}
    }

    \captionsetup{justification=raggedright,singlelinecheck=false}
    \caption{Plot of the order parameter $\xi_*$ in the $(\kappa, \tau)$ plane for (a) $\gamma= 10^{-2}$ and (b) $\gamma=10^{-1}$. The solid and dashed white lines are the true critical line $\tau =1$ and crossover line $\tau=\kappa$, respectively. Other parameters used are $d=30$, $\lambda=10^{-5}$, and $\sigma=0$. A small shift $\xi_* \rightarrow \xi_* + 10^{-5}$ was made to avoid numerical instability of the logarithm at small $\xi_*$. Depression of the order parameter in the pseudogap region is highly visible for very small value of $\gamma$ used in (a). (c) Comparison of the order parameter extracted from Eq.~\eqref{eq:numerical-xi-fit} (hollow marks) with the theoretical curves obtained by solving the self-consistency equation \eqref{SC-xi-again} at $\kappa=0.6$.}
    \label{fig:strict_phase_diagram_d30}
\end{figure}

Figure \ref{fig:strict_phase_diagram_d30} shows the map of $\xi_*$ extracted from this scheme for $\gamma = 10^{-1}$ and $\gamma=10^{-2}$. In the latter case, the pseudogap behavior with suppressed order parameter is prominent over the region bounded by $\tau=1$ and $\tau=\kappa$ in accordance with the predictions of the Landau theory. An additional comparison between $\xi_*$ extracted from the empirical scheme in Eq.~\eqref{eq:numerical-xi-fit} and the self-consistency equation \eqref{SC-xi-again} finds an excellent fit, as shown in Fig.~\ref{fig:strict_phase_diagram_d30}(c), bolstering the reliability of the extraction scheme. 

\begin{figure}[]
    \centering
    \hspace{-0.8cm}
    \includegraphics[width=0.9\columnwidth]{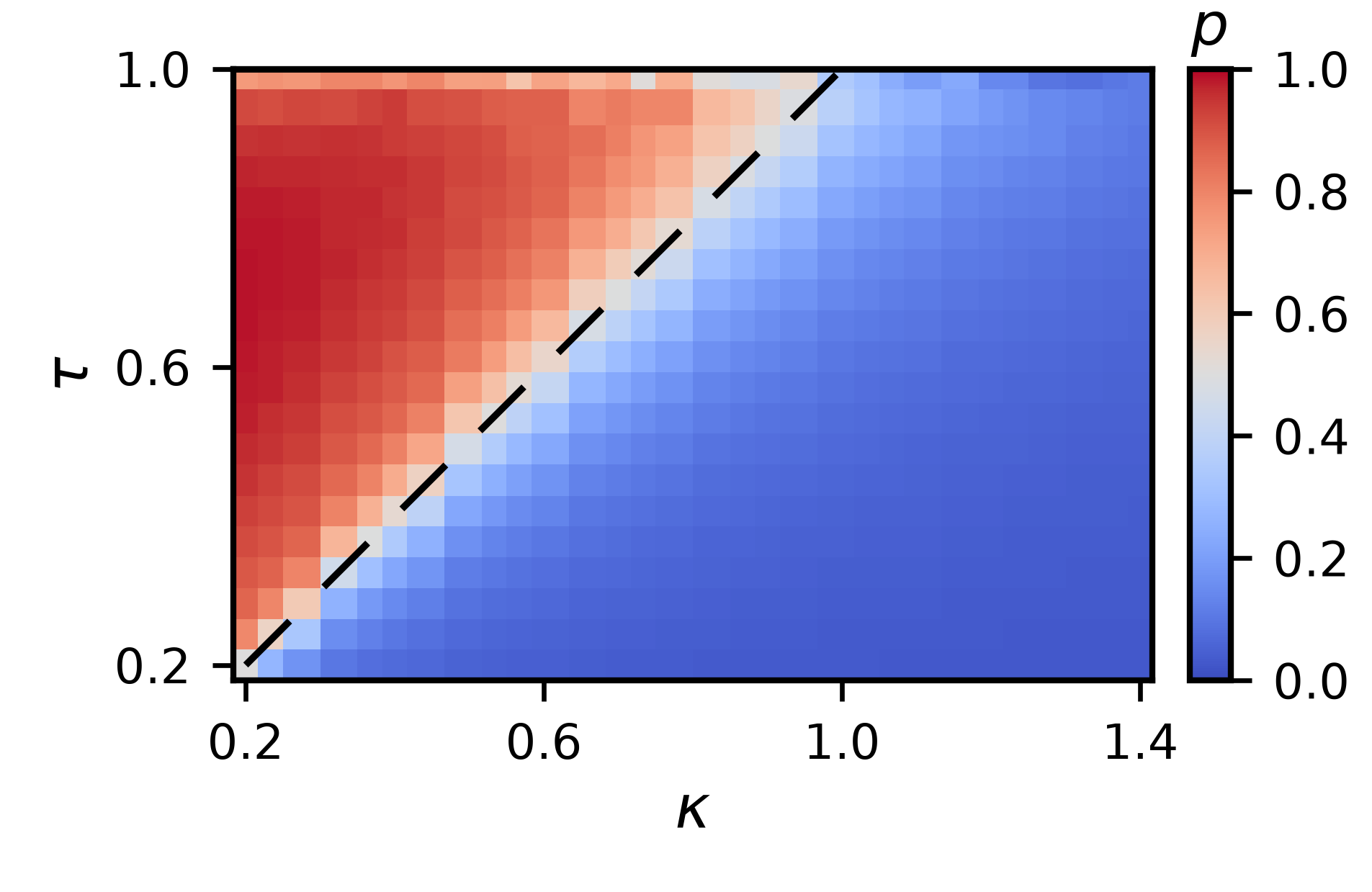}
    \caption{Map of the numerically extracted power-law exponents $p$ in the assumed scaling relation $\xi_* \propto \gamma^p$. The ordered phase $\tau<1$ separates into two distinct scaling regimes with $p\approx 0$ and $p \approx 1$ (pseudogap region), divided by the theoretically predicted crossover line $\tau = \kappa$ (dashed line). }
    \label{fig:gamm_scaling}
\end{figure}

We also performed calculations in which $\xi_*$ obtained from the empirical formula \eqref{eq:numerical-xi-fit} over various values of $(\gamma, \kappa, \tau)$ were fit to the putative power-law relation $\xi_* \propto \gamma^{p(\kappa, \tau)}$. To extract the exponent $p(\kappa, \tau)$ at each fixed $(\kappa,\tau)$, we fit $\log \xi_*$ against $\log \gamma$ using $\gamma\in\{10^{-2.5},10^{-2},10^{-1.5},10^{-1}\}$, and identified $p(\kappa, \tau)$ with the obtained slope. The resulting map in Fig. \ref{fig:gamm_scaling} in the $(\kappa, \tau)$ plane shows a rapid but smooth increase from $p(\kappa, \tau) \approx 0$ to $p(\kappa, \tau) \approx 1$ at the crossover line $\tau =\kappa$, again in good agreement with the predictions of the Landau theory.

\section{Summary and discussion}
\label{sec:discussion}
We have demonstrated that the interpolation singularity of linear in-context learning can be interpreted as a genuine critical phenomenon of a quenched disordered system. The key diagnostic is the contrast between annealed and quenched averaging. By performing both calculations and contrasting their outcome, we conclude that the double-descent peak is indeed a manifestation of a divergent connected fluctuation in the learning parameter.

Integrating the self-consistent equation for the renormalized ridge parameter immediately leads to the Landau potential $V(\xi)$. Minimizing the Landau potential leads to the physical order parameter $\xi_*$. The normalized sample complexity $\tau$ and the bare ridge parameter $\lambda$ are mapped to temperature and magnetic-field variables in the Landau theory, respectively. The associated Landau susceptibility, proportional to the inverse of the curvature $V'' (\xi_*)$, controls the singular fluctuation contribution to the quenched test error and explains its divergence as the divergent susceptibility at the critical point in the usual sense of critical phenomena. The meaning of the order parameter is, being closely related to the density of zero modes in the relaxation matrix, geometric rather than tied to the conventional notion of spontaneous symmetry breaking. In line with such interpretation, the Landau potential generically contains a cubic term in the order parameter that leads to critical exponents $(\beta_{\rm cr},\delta_{\rm cr},\gamma_{\rm cr})=(1,2,1)$.

We identify a pseudogap-like behavior for large context lengths. This occurs above the crossover temperature $\tau = \kappa$ and below the true critical temperature $\tau =1$, when the task diversity parameter $\kappa$ is less than one. A substantial depletion of the order parameter proportional to $\gamma$ characterizes the pseudogap region. The endpoint $\kappa=1$ provides a particularly intriguing situation where, in the strict $\gamma=0$ limit, the Landau functional acquires a nonanalytic $\xi^{5/2}$ dependence with the associated critical exponents $(\beta_{\rm cr},\delta_{\rm cr},\gamma_{\rm cr})=(2,3/2,1)$. 

The order parameter $\xi_*$ can be extracted directly from numerical solutions of the original learning problem, without relying on the cavity self-consistency equation. The results are in very good agreement with conclusions drawn from analysis of the Landau theory. The agreement supports the idea that there are indeed aspects of in-context learning that can be understood in the framework of conventional Landau theory. The treatment of input vectors as random variables in the learning model suggests close analogy to the disorder problem in physics. In the future, we plan to employ powerful techniques of disorder such as the replica method to bring out heightened connection between the theory of learning and the theory of disordered systems. 


\acknowledgments JHH was supported by the National Research Foundation of Korea (NRF) grant funded by the Korea government(MSIT) (Grant No. 2023R1A2C1002644 and No. RS-2024-00410027). He thanks Yue M. Yu for sharing his insights on linear ICL.

\bibliography{LICL-PRXI}

\appendix

\section{Derivation of the formula for $\hat y_{\ell+1}$ in linear in-context learning}
\label{app:how-to-derive-y-ell+1} 

The ML network is provided with a set of $\ell$ input-output pairs $\{ x_m \}_{m=1}^\ell$, $x_m=({\bf x}_m,y_m)$, whose relation must be inferred to return a reliable prediction for the output $y_{\ell+1}$ to a new input $\x_{\ell+1}$. Schematically, the task at hand for ICL is to correctly capture the map 
\begin{align} {\bf x}_{\ell+1} \xrightarrow{\rm ICL} y_{\ell+1} . 
\end{align} 

In the theory of LICL~\cite{pehlevan25}, this is done by appealing to the linear self-attention ($\ell$SA) mechanism~\cite{ahn24}: 
\begin{align}
x'_n = x_n + \frac{1}{\ell} \sum_{m=1}^{\ell} (q_n \cdot k_m ) v_m . 
\label{linear-SA} 
\end{align}
where $n$ runs from 1 through $\ell+1$. We set $x_{\ell+1}=({\bf x}_{\ell+1},0)$ with 0 being the placeholder for $y_{\ell+1}$. The prediction for $y_{\ell+1}$, denoted $\hat y_{\ell+1}$, is then obtained as the $(d+1)$-th component of $x'_{\ell+1}$. The key, query, and value vectors are defined in the usual way of the Transformer:
\begin{align} 
    k_m = K x_m , ~ q_m = Q x_m , ~ v_m = V x_m  ,
\end{align}
with suitable choices of matrices $K, Q, V$. 
The value vector $v_m \in \mathbb{R}^{d+1}$ has the same dimension as the token vector $x_m$, but the key and query vector dimensions may be different. In reality, the inner product
\begin{align}
    q_n \cdot k_m = x_n^\top (Q^\top K) x_m \equiv x_n^\top M x_m 
\end{align}
depends on a single matrix $M = Q^\top K$, not on $Q$ and $K$ individually. This is also the case with the full Transformer architecture where, instead of the linearized (and un-normalized) weight $q_n \cdot k_m$ weight, a full softmax function is employed to enforce the SA mechanism. The adoption of $\ell$SA layer allows for the development of an elegant and analytically tractable theory of ICL as described below. 

Using $M=Q^\top K$, the prediction $\hat y_{\ell+1}$ follows from the $\ell$SA formula \eqref{linear-SA} as 
\begin{align}
\hat y_{\ell+1} =  x'_{\ell+1,d+1}  = \frac{1}{\ell} \sum_{m=1}^{\ell} (x_{\ell+1}^\top M x_m ) [ V x_m ]_{d+1} . 
\label{y-l-plus-1} 
\end{align}
%
Due to the break-up of each token $x_m$ into a $d$-dimensional input component ${\bf x}_m$ and a scalar output component $y_m$, it proves convenient to make an analogous decomposition of the $M$ and $V$ matrices as
\begin{align} 
M=\begin{pmatrix} {\bm M}_{xx} & {\bf M}_{xy}\\ {\bf M}_{yx}^\top & M_{yy}\end{pmatrix},\quad
V=\begin{pmatrix} {\bm V}_{xx} & {\bf V}_{xy}\\ {\bf V}_{yx}^\top & V_{yy}\end{pmatrix},
\end{align}
with ${\bm M}_{xx}, {\bm V}_{xx}\in\mathbb{R}^{d\times d}$, ${\bf M}_{xy},{\bf M}_{yx},{\bf V}_{xy},{\bf V}_{yx}\in\mathbb{R}^{d}$, and $M_{yy},V_{yy}\in\mathbb{R}$. Since $x_{\ell+1} = (\x_{\ell+1}~ 0)^\top$, we get 
\begin{align}
x_{\ell+1}^\top M x_m &=\x_{\ell+1}^\top {\bm M}_{xx} \x_m + y_m  \x_{\ell+1}^\top {\bf M}_{xy} , 
\label{x-ell+1} 
\end{align} 
where ${\bf M}_{yx}, M_{yy}$ do not appear and may as well be set to zero from the outset. Also in 
\begin{align}
\left[ V x_m \right]_{d+1} & ={\bf V}_{yx}^\top \x_m  + V_{yy} y_m , 
\label{how-to-define-M-and-Z} 
\end{align}
we choose to set ${\bf V}_{yx}=0$ and focus on the part containing $y_m$. This assumption allows us to write $$\left[ V x_m \right]_{d+1} \rightarrow V_{yy} y_m , $$ and furthermore let $V_{yy}$ be absorbed in the re-definition of ${\bm M}_{xx}$ and ${\bf M}_{xy}$ in \eqref{x-ell+1}. As a result, we obtain a very simple expression for $\hat y_{\ell+1}$ based on $\ell$SA mechanism: 
\begin{align}
\hat y_{\ell+1} & 
 = \frac{1}{\ell} \sum_{m=1}^\ell y_m \x_{\ell+1}^\top ( {\bm M}_{xx} \x_m + y_m {\bf M}_{xy} ) \nn 
& = \frac{1}{\ell} \sum_{m=1}^\ell (q_{\ell+1} \cdot  k_m ) y_m 
\label{eq:yhat_reduced_same}
\end{align}
where $q_{\ell+1} = Q x_{\ell+1}$ and $k_m = K x_m$. It allows us to treat $\hat y_{\ell+1}$ as the average of existing labels $y_m$, weighted by the overlap between the query vector $q_{\ell+1}$ and the key vector $k_m$ for each $m$-th token. The formula \eqref{eq:yhat_reduced_same} reproduces the result in \cite{pehlevan25}, using the formalism and notations most familiar to physicists. 

For further progress, we re-cast the inner product $q_{\ell+1} \cdot k_m$ as 
\begin{align}
     q_{\ell+1} \cdot k_m & = {\bf x}^\top_{\ell+1}\begin{pmatrix} {\bm M}_{xx} & {\bf M}_{xy} \end{pmatrix} x_m \nn 
    & = \Bigl( \begin{pmatrix} {\bm M}_{xx} & {\bf M}_{xy} \end{pmatrix},{\bf x}_{\ell+1}x_m^\top \Bigr)_{\rm F} \, , 
\end{align}
where the last expression is the Frobenius product between two $d\times (d+1)$-dimensional matrices $\begin{pmatrix} {\bm M}_{xx} & {\bf M}_{xy} \end{pmatrix}$ and ${\bf x}_{\ell+1} x_m^\top$. Now, $\hat y_{\ell+1}$ is given by the Frobenius product 
\begin{align}
 \hat y_{\ell+1} & =  \left(\begin{pmatrix} {\bm M}_{xx} & {\bf M}_{xy} \end{pmatrix},  \frac{1}{\ell} \sum_{m=1}^\ell y_m  {\bf x}_{\ell+1} x_m^\top \right)_{\rm F} \,  . 
\end{align}
A more intuitive expression for $y_{\ell+1}$ is achieved by {\it vectorizing} the two matrices:  
\begin{align} 
{\bf P} & = {\rm vec} \left[ \begin{pmatrix} {\bm M}_{xx}/d & {\bf M}_{xy}\end{pmatrix} \right] \, , \nn
{\bf H} & = {\rm vec} \left[ \frac{1}{\ell} \sum_{m=1}^\ell y_m  {\bf x}_{\ell+1}  ( \widetilde x_m )^\top \right]  \, , \nn 
\widetilde x_m
    & \equiv
    \begin{pmatrix}
        d\,\mathbf{x}_m \\
        y_m
    \end{pmatrix} \, . 
\label{general-P-and-H-defined} 
\end{align} 
The re-scaling of the spatial component of $x_m$ by $d$ is intended to keep all the terms in ${\bf H}$ scale similarly in the large-$d$ limit. Using the vectorization convention ${\rm vec}[{\bf u}v^\top] = {\bf u}\otimes v$, the vector $\bf H$ can also be written as in Eq.~\eqref{general-H}. Now the prediction for $y_{\ell+1}$ becomes neatly formulated as an inner product of two vectors:
\begin{align}
\hat y_{\ell+1}  = {\bf P} \cdot {\bf H} . 
\label{y-as-linear-response} 
\end{align}
One of them, called the learning vector ${\bf P}$, consists of $d(d+1)$ parameters to be tuned by machine learning. The other, denoted ${\bf H}$, contains all the existing information on the context $\{ x_m \}_{m=1}^\ell$, plus the information on the test input $\x_{\ell+1}$ needed to accurately predict $y_{\ell+1}$.  The purpose of training (learning) is to identify the optimal learning vector ${\bf P}$. One may well regard the formula \eqref{y-as-linear-response} together with the ${\bf P}$ and ${\bf H}$ vectors in \eqref{general-P-and-H-defined} as the {\it definition} of the LICL scheme~\cite{pehlevan25}.

\section{Computation of the annealed relaxation matrix and source vector}
\label{app:annealed-R-and-S}
We motivate the annealing approach by decomposing the field vector ${\bf H}^\mu$ in \eqref{H-mu} as the average plus the fluctuation:
\begin{align} 
{\bf H}^\mu  & = {\bf x}^\mu_{\ell+1} \otimes \left(  \frac{1}{\ell} \sum_{m=1}^\ell y_m^\mu  \widetilde x_m^\mu \right) \nn 
& = {\bf x}^\mu_{\ell+1} \otimes \left( v^\mu + \frac{1}{\sqrt{\ell}} \eta^\mu \right) .
\label{H-mean-plus-fluc} 
\end{align} 
The $v^\mu$ in the second line is the conditional average over ${\bf x}^\mu_m$ and $\epsilon^\mu_m$ for a fixed ${\bf w}^\mu$:
\begin{align} v^\mu  & =  \frac{1}{\ell} \sum_{m=1}^\ell \mathbb{E}[y_m^\mu  \widetilde x_m^\mu|{\bf w}^\mu] =   \begin{pmatrix} {\bf w}^\mu \\ a^\mu \end{pmatrix} \, , \nn a^\mu & = \frac{|\mathbf w^\mu|^2}{d}+\sigma^2 . 
\end{align} 
In accordance with \cite{pehlevan25}, we assume ${\bf w}^\mu \sim \mathsf{Unif}(S^{d-1} (\sqrt{d}))$---which is indistinguishable from ${\cal N}(0,I_d)$ for sufficiently large $d$--- and set $|{\bf w}^\mu |^2 = d, \, a^\mu = 1 + \sigma^2$. The fluctuation is captured by $\eta^\mu$: 
\begin{align} 
\eta^\mu = \sqrt{\ell} \left( \frac{1}{\ell} \sum_{m=1}^\ell y_m^\mu \widetilde{x}_m^\mu - v^\mu \right) , 
\end{align} 
where the $\sqrt{\ell}$ factor reflects the deviation from the mean $\propto \sqrt{\ell}$. The source vector ${\bf S}$ in \eqref{R-and-S} is similarly decomposed:
\begin{align}
{\bf S}  = \frac{d}{n}\sum_\mu y^\mu_{\ell+1} {\bf x}^\mu_{\ell+1} \otimes \left( v^\mu + \frac{1}{\sqrt{\ell}} \eta^\mu \right) . 
\end{align}

Inserting ${\bf H}^\mu$ in \eqref{H-mean-plus-fluc} into ${\bf H}^\mu ({\bf H}^\mu )^\top$ gives the conditional average
\begin{align}
\mathbb{E}\big[{\bf H}^\mu ({\bf H}^\mu )^\top|{\bf w}^\mu,{\bf x}^\mu_{\ell+1}\big] & =  {\bf x}^\mu_{\ell+1} ( {\bf x}^\mu_{\ell+1} )^\top \otimes \left( v^\mu (v^\mu )^\top  + \frac{d}{\ell}  \Sigma^\mu \right) . 
\label{HHt}
\end{align}
The self-energy part $\Sigma^\mu$ is given by
\begin{align}
\mathbb{E}[\eta^\mu (\eta^\mu)^\top ] 
 = \begin{pmatrix} d a^\mu I_d + {\bf w}^\mu ({\bf w}^\mu )^\top & 2 a^\mu {\bf w}^\mu \\ 2 a^\mu ({\bf w}^\mu )^\top & 2 (a^\mu )^2  \end{pmatrix} \equiv d \cdot \Sigma^\mu  .  
\label{79}
\end{align} 
%
%
Performing the average over $\x_{\ell+1}^\mu$ yields
\begin{align} 
\mathbb{E}\big[ {\bf x}^\mu_{\ell+1} ( {\bf x}^\mu_{\ell+1} )^\top \big] = \frac{1}{d} I_d , ~~ 
\mathbb{E}\big[ y_{\ell+1}^\mu {\bf x}_{\ell+1}^\mu |{\bf w}^\mu \big] =  \frac{1}{d} {\bf w}^\mu .
\end{align} 
The expectation values of the relaxation matrix and the source vector, conditioned on the task vectors $\{{\bf w}^\mu\}_{\mu=1}^n$, are then obtained as
\begin{align}
\mathbb{E}[{\bm R}|\{{\bf w}^\mu\}] =& I_d \otimes (E + \lambda I_{d+1}) \, ,  \nn 
\mathbb{E}[{\bf S}|\{{\bf w}^\mu\}] =& \frac{1}{n} \sum_{\mu=1}^n {\bf w}^\mu  \otimes v^\mu , 
\label{R-S_annealed_1}
\end{align}
with
\begin{align}
    E = \frac{1}{n} \sum_{\mu=1}^n \left[   v^\mu (v^\mu )^\top  + \frac{d}{\ell}  \Sigma^\mu  \right] . 
\end{align}
%
%
The matrix $E$ can be reorganized as
\begin{align}
       E = \frac{1}{n}\sum_{\mu=1}^n \big[E({\bf w}^\mu) + \delta E({\bf w}^\mu) \big],
\end{align}
where
\begin{gather}
    E({\bf w}) = \begin{pmatrix}
        \gamma I_d+{\bf w}{\bf w}^\top & (1+\sigma^2) {\bf w} \\
        (1+\sigma^2) {\bf w}^\top & (1+\sigma^2)^2
    \end{pmatrix}, \nn [3pt]
    {\delta E}({\bf w})
    =
    \frac{1}{\ell}
    \begin{pmatrix}
    {\bf w}{\bf w}^\top & 2(1+\sigma^2){\bf w}
    \\
    2(1+\sigma^2){\bf w}^\top & 2(1+\sigma^2)^2
    \end{pmatrix} . 
\label{Ew}
\end{gather}
As shorthand, we defined
\begin{align}
     \gamma \equiv \frac{1+\sigma^2}{\alpha} , 
\label{gamma}
\end{align}
where $\alpha=\ell/d$ is the normalized context length. In the joint asymptotic limit, $\delta E({\bf w}^\mu)$ vanishes and we can write
\begin{align}
    E\xrightarrow{d\rightarrow\infty}&\frac{1}{n}\sum_{\mu=1}^n E({\bf w}^\mu) \nn
    &=  \begin{pmatrix}
        \gamma I_d+{\bm C}^{\rm tr} & (1+\sigma^2) {\bf b}^{\rm tr} \\
        (1+\sigma^2) ({\bf b}^{\rm tr})^\top & (1+\sigma^2)^2
    \end{pmatrix} \nn
    &\equiv E^{\rm tr}
\label{E_train-app}
\end{align}
with the mean and covariance matrix of the $k$ unique training task vectors ${\bf w}_1,\dots,{\bf w}_k$ in Eq.~\eqref{w_train},
\begin{align}
    {\bf b}^{\rm tr}
    \equiv
    \frac{1}{k}\sum_{p=1}^k {\bf w}_p,
    \qquad
    {\bm C}^{\rm tr}
    \equiv
    \frac{1}{k}\sum_{p=1}^k {\bf w}_p{\bf w}_p^\top . 
\end{align}
Here we assumed that the number of training samples $n$ far exceeds the task diversity $k$, such that we can replace the average over $n$ samples of task vectors by an average over $\mathsf{Unif}(\Omega)$:
\begin{align}
    \frac{1}{n}\sum_\mu E({\bf w}^\mu) \rightarrow \frac{1}{k}\sum_p E({\bf w}_p) .
\label{n-to-k}
\end{align}
For the same reason, in Eq.~\eqref{R-S_annealed_1} we can make the replacement
\begin{align}
    \frac{1}{n}\sum_\mu {\bf w}^\mu \otimes v^\mu \rightarrow \frac{1}{k}\sum_p {\bf w}_p \otimes \begin{pmatrix} {\bf w}_p \\ 1+\sigma^2 \end{pmatrix}.
\end{align}
This completes the annealing of the relaxation matrix and the source vector:
\begin{align}
    \mathbb{E}_{\rm tr}[{\bm R}] &= I_d \otimes \big( E^{\rm tr} + \lambda I_{d+1} \big), \nn
    \mathbb{E}_{\rm tr}[{\bf S}] &= \frac{1}{k}\sum_{p=1}^k {\bf w}_p \otimes \begin{pmatrix} {\bf w}_p \\ 1+\sigma^2\end{pmatrix}.
\end{align}

\section{Marchenko-Pastur distribution}
\label{app:Wishart-and-MP}

The eigenvalues of the Wishart matrix 
\begin{align}
    \frac{1}{k} \sum_{p=1}^k {\bf w}_p {\bf w}_p^\top , ~~ {\bf w}_p \sim {\cal N} (0, I_d ) ,
\end{align}
follow the MP distribution 
\begin{align} 
\rho_{\rm MP} (\lambda) = \frac{\sqrt{(\lambda_+ - \lambda) (\lambda- \lambda_- )}}{2\pi \lambda/\kappa}, ~~ \lambda_\pm = (1\pm \kappa^{-1/2} )^2 ,  
\end{align} 
when $k$ and $d$ are taken to infinity simultaneously with the fixed ratio $\kappa = k/d$ \cite{bai10}. The integral $\int_{\lambda_-}^{\lambda_+} \rho_{\rm MP} (\lambda) d\lambda = 1$ saturates to one, meaning that all eigenvalues are, statistically speaking, nonzero. The result holds when the task diversity $k$ exceeds the dimension of the task  vector $d$. The Wishart matrix reaches full rank with its $d$ eigenvalues distributed with probability $\rho_{\rm MP}(\lambda)$. 

In the other case $\kappa=k/d < 1$, the Wishart matrix is low-rank with some eigenvalues equal to zero. The eigenvalue distribution follows
\begin{align} 
\rho_{\rm MP} (\lambda) & = \frac{\sqrt{(\lambda_+ - \lambda) (\lambda- \lambda_- )}}{2\pi \lambda/\kappa} + \left( 1- \kappa \right) \delta (\lambda) , \nn 
\lambda_\pm & = (\kappa^{-1/2} \pm 1 )^2 . 
\end{align} 
The continuous part of the distribution gives $\int_{\lambda_-}^{\lambda_+} d\lambda \rho_{\rm MP} (\lambda) = \kappa$. The delta function gives the remaining weight $1-\kappa = (d-k)/d$. There are $\approx d-k$ zero eigenvalues, and only $k$ nonzero eigenvalues. 


Throughout this work, the MP law is applied via its Stieltjes transform~\cite{bai10},
\begin{align}
    I_\kappa(z)&=\int {\rm d}\lambda'\frac{\rho_{\rm MP}(\lambda')}{\lambda'+z}
    \nn
    &=\frac{2}{z+1-1/\kappa+\sqrt{(z+1-1/\kappa)^2+4z/\kappa}
    },\quad z>0.
    \label{Stieltjes-app}
\end{align}

\section{General testing error under annealing}
\label{app:test-error-under-A}

Here we derive the general annealed error in Eq.~\eqref{annealed_test_error_general_b} starting from Eq.~\eqref{annealed_test_error},
\begin{align}
    {\cal E}^A = & \mathbb{E}_{\rm te}[(y^{\rm te}_{\ell+1})^2] -2{\bf P}^A_\infty \cdot \mathbb{E}_{\rm te} [y^{\rm te}_{\ell+1}{\bf H}^{\rm te}] \nn
    & +({\bf P}^A_\infty)^\top\mathbb{E}_{\rm te} [{\bf H}^{\rm te}({\bf H}^{\rm te})^\top]{\bf P}^A_\infty  . 
\label{annealed_test_error-app}
\end{align}
The first two averages are evaluated as
\begin{align}
    \mathbb{E}_{\rm te} [(y^{\rm te}_{\ell+1})^2] =& 1 + \sigma^2  , \nn 
    \mathbb{E}_{\rm te} [y^{\rm te}_{\ell+1}{\bf H}^{\rm te}] =& \frac{1}{d}\mathbb{E}_{{\bf w}^{\rm te}\sim{\cal P}^{\rm te}}\left[ {\bf w}^{\rm te} \otimes v^{\rm te} \right] \nn =& \frac{1}{d} {\rm vec}\left[\begin{pmatrix} {\bm C}^{\rm te} & (1+\sigma^2){\bf b}^{\rm te}\end{pmatrix}\right] .  
\end{align}
For the third average in Eq.~\eqref{annealed_test_error-app}, note that ${\bf H}^{\rm te}$ in Eq.~\eqref{Ht} has the same form as ${\bf H}^\mu$ in Eq.~\eqref{H-mu} except for the change in superscript $\mu\rightarrow {\rm te}$. Following the steps similar to Eqs. \eqref{HHt} to \eqref{Ew}, we obtain
\begin{align}
    \mathbb{E}_{\rm te} [{\bf H}^{\rm te}({\bf H}^{\rm te})^\top] =& \frac{1}{d}I_d \otimes \mathbb{E}_{{\bf w}^{\rm te}\sim{\cal P}^{\rm te}}\left[ E({\bf w}^{\rm te}) + \delta  E({\bf w}^{\rm te}) \right], 
\end{align}
where $E({\bf w}^{\rm te})$ and $\delta E({\bf w}^{\rm te})$ are as defined in Eq.~\eqref{Ew}. $\delta E({\bf w}^{\rm te})$ does not cause any divergence for nonzero $\ell$ and vanishes in the joint asymptotic limit, so that we can write
\begin{align}
    & \mkern-21mu \mathbb{E}_{\rm te} [{\bf H}^{\rm te}({\bf H}^{\rm te})^\top] \xrightarrow{d\rightarrow\infty} \frac{1}{d} I_d \otimes E^{\rm te}, \nn[3pt]
    E^{\rm te} & \equiv \mathbb{E}_{{\bf w}^{\rm te}\sim{\cal P}^{\rm te}}\left[ E({\bf w}^{\rm te})\right] \nn
    & = \begin{pmatrix}
        \gamma I_d+{\bm C}^{\rm te} & (1+\sigma^2) {\bf b}^{\rm te} \\
        (1+\sigma^2) ({\bf b}^{\rm te})^\top & (1+\sigma^2)^2
    \end{pmatrix} .
\label{Ete}
\end{align}
Combining the above results with the definition of ${\bf P}^A_\infty$ in Eq.~\eqref{PAinfty} yields
\begin{widetext}
\begin{align}
    {\bf P}^A_\infty \cdot \mathbb{E}_{\rm te} [y^{\rm te}_{\ell+1}{\bf H}^{\rm te}]
    =& {\rm vec}\left[
    \begin{pmatrix}
        {\bm C}^{\rm tr} & (1+\sigma^2){\bf b}^{\rm tr}
    \end{pmatrix}
    (E^{\rm tr}+\lambda I_{d+1})^{-1}\right] \cdot \frac{1}{d} {\rm vec}\left[\begin{pmatrix} {\bm C}^{\rm te} & (1+\sigma^2){\bf b}^{\rm te}\end{pmatrix}\right] \nn
    =& \frac{1}{d}{\rm Tr} \left[ \begin{pmatrix} {\bm C}^{\rm tr} & (1+\sigma^2){\bf b}^{\rm tr} \end{pmatrix}
    (E^{\rm tr}+\lambda I_{d+1})^{-1}
    \begin{pmatrix} {\bm C}^{\rm te} & (1+\sigma^2){\bf b}^{\rm te} \end{pmatrix}^{\top} \right], \nn
    ({\bf P}^A_\infty)^\top\mathbb{E}_{\rm te} [{\bf H}^{\rm te}({\bf H}^{\rm te})^\top]{\bf P}^A_\infty 
    =& {\rm vec}\left[
    \begin{pmatrix}
        {\bm C}^{\rm tr} & (1+\sigma^2){\bf b}^{\rm tr}
    \end{pmatrix}
    (E^{\rm tr}+\lambda I_{d+1})^{-1}\right]^\top 
    \frac{1}{d}(I_d \otimes E^{\rm te}) \, {\rm vec}\left[
    \begin{pmatrix}
        {\bm C}^{\rm tr} & (1+\sigma^2){\bf b}^{\rm tr}
    \end{pmatrix}
    (E^{\rm tr}+\lambda I_{d+1})^{-1}\right] \nn
    =& \frac{1}{d}{\rm Tr} \left[ \begin{pmatrix} {\bm C}^{\rm tr} & (1+\sigma^2){\bf b}^{\rm tr} \end{pmatrix}
    (E^{\rm tr}+\lambda I_{d+1})^{-1} E^{\rm te} (E^{\rm tr}+\lambda I_{d+1})^{-1}
    \begin{pmatrix} {\bm C}^{\rm tr} & (1+\sigma^2){\bf b}^{\rm tr} \end{pmatrix}^{\top}
    \right] ,
\end{align}
and consequently the annealed error formula \eqref{annealed_test_error_general_b} for a general test distribution ${\cal P}^{\rm te}$. 

One can show that terms containing ${\bf b}^{\rm tr}$ or ${\bf b}^{\rm te}$ can be neglected within the scope of this work (see App.~\ref{app:btr=0}). Eq.~\eqref{annealed_test_error_general_b} then reduces to
\begin{align}
    {\cal E}^A = 1+\sigma^2 - \frac{2}{d} {\rm Tr}\left[{\bm C}^{\rm tr} ({\bm C}^{\rm tr}+\widetilde\lambda I_d)^{-1}{\bm C}^{\rm te}\right]
    + \frac{1}{d}{\rm Tr}\left[{\bm C}^{\rm tr} ({\bm C}^{\rm tr}+\widetilde\lambda I_d)^{-1}({\bm C}^{\rm te}+\gamma I_d)({\bm C}^{\rm tr}+\widetilde\lambda I_d)^{-1}{\bm C}^{\rm tr}\right].
\label{annealed_test_error_general-app}
\end{align}
\end{widetext}
with $\widetilde\lambda = \gamma + \lambda$. We can simplify Eq.~\eqref{annealed_test_error_general-app} by introducing
\begin{align}
    K\equiv{\bm C}^{\rm tr}\left({\bm C}^{\rm tr}+\widetilde\lambda I_d\right)^{-1}.
\end{align}
Then Eq.~\eqref{annealed_test_error_general-app} becomes
\begin{align}
    {\cal E}^A=1+\sigma^2+\frac{1}{d}{\rm Tr}\left[{\bm C}^{\rm te}(K^2-2K)\right]+\frac{\gamma}{d}{\rm Tr}\left[K^2\right].
\end{align}
Using the identity
\begin{align}
    K^2-2K&=(K-I_d)^2-I_d  \nn
    &=\widetilde\lambda^2\left({\bm C}^{\rm tr}+\widetilde\lambda I_d\right)^{-2}-I_d,
\end{align}
and assuming that the test task distribution is normalized as
\begin{align}
    \frac{1}{d}{\rm Tr}\left[{\bm C}^{\rm te}\right]=1,
\end{align}
which is asymptotically exact for the ICL and IDG tests, the expression further reduces to
\begin{align}
    {\cal E}^A=\sigma^2+\frac{1}{d}{\rm Tr}\left[\left(\widetilde\lambda^2{\bm C}^{\rm te}+\gamma({\bm C}^{\rm tr})^2\right)\left({\bm C}^{\rm tr}+\widetilde\lambda I_d\right)^{-2}\right].
\label{annealed_test_error_simple}
\end{align}
For later convenience, we separate the distribution mismatch ${\bm C}^{\rm te}-{\bm C}^{\rm tr}$ and the ${\bm C}^{\rm te}$-independent terms:
\begin{align}
    {\cal E}^A =& \frac{\widetilde\lambda^2}{d}{\rm Tr}\left[\left({\bm C}^{\rm te}-{\bm C}^{\rm tr}\right)\left({\bm C}^{\rm tr}+\widetilde\lambda I_d\right)^{-2}\right] \nn
    &+\sigma^2 + \frac{\widetilde\lambda}{d}{\rm Tr}\left[{\bm C}^{\rm tr}\left({\bm C}^{\rm tr}+\widetilde\lambda I_d\right)^{-1}\right] \nn
    &- \frac{\lambda}{d}{\rm Tr}\left[({\bm C}^{\rm tr})^2\left({\bm C}^{\rm tr}+\widetilde\lambda I_d\right)^{-2}\right] .
\end{align}

\section{Order estimation of the ${\bf b}^{\rm tr}$-dependent terms in the annealed test error}
\label{app:btr=0}

To show that the ${\bf b}^{\rm tr}$-dependent terms in the annealed test error [Eq. \eqref{annealed_test_error_general_b}] are negligible in the joint asymptotic limit, we first compute the elements of the inverse matrix $(E^{\rm tr}+\lambda I_{d+1})^{-1}$ explicitly,
\begin{align}
    (E^{\rm tr}+\lambda I_{d+1})^{-1} = \begin{pmatrix}
        {\bm D} & {\bf g} \\
        {\bf g}^\top & q
    \end{pmatrix},
\end{align}
with
\begin{align}
    \frac{1}{q} =& \lambda+(1+\sigma^2)^2\left[1-({\bf b}^{\rm tr})^\top ({\bm C}^{\rm tr}+\widetilde\lambda I_d)^{-1}{\bf b}^{\rm tr} \right], \nn
    {\bf g} =& -q(1+\sigma^2)({\bm C}^{\rm tr}+\widetilde\lambda I_d)^{-1}{\bf b}^{\rm tr}, \nn
    {\bm D} =& ({\bm C}^{\rm tr}+\widetilde\lambda I_d)^{-1}+\frac{1}{q}{\bf g}{\bf g}^
    \top.
\end{align}
The ${\bf b}^{\rm tr}$-dependent terms of the annealed test error ${\cal E}^A$ can be summarized as
\begin{widetext}
\begin{align}
    \Delta {\cal E}^A \equiv &\frac{\widetilde\lambda}{d} \bigg[\frac{1}{q}{\bf g}^\top \big(\widetilde\lambda({\bm C}^{\rm tr}+\widetilde\lambda I_d)^{-1}{\bm C}^{\rm te}-\gamma{\bm C}^{\rm tr} ({\bm C}^{\rm tr}+\widetilde\lambda I_d)^{-1}\big){\bf g}
    +2\widetilde\lambda(1+\sigma^2){\bf g}^\top ({\bm C}^{\rm tr}+\widetilde\lambda I_d)^{-1} {\bf b}^{\rm te} \nn
    &\qquad +\frac{\widetilde\lambda}{q}\lVert{\bf g}\rVert^2\Big(\frac{1}{q}{\bf g}^\top \big({\bm C}^{\rm te}+\gamma I_d\big){\bf g}+2(1+\sigma^2){\bf g}^\top{\bf b}^{\rm te}+q(1+\sigma^2)^2\Big)\bigg].
\end{align}
In case of the ICL setting (${\bf b}^{\rm te}=0,\,{\bm C}^{\rm te}=I_d$) this reduces to
\begin{align}
    \Delta {\cal E}^A_{\rm ICL} =& 
    \frac{q\widetilde\lambda(1+\sigma^2)^2}{d}\bigg[2\widetilde\lambda(\gamma+1)({\bf b}^{\rm tr})^\top({\bm C}^{\rm tr}+\widetilde\lambda I_d)^{-3}{\bf b}^{\rm tr}
    - 2\gamma\lVert({\bm C}^{\rm tr}+\widetilde\lambda I_d)^{-1}{\bf b}^{\rm tr}\rVert^2 \nn
    & \qquad \qquad \qquad + q\widetilde\lambda(1+\sigma^2)^2\lVert({\bm C}^{\rm tr}+\widetilde\lambda I_d)^{-1}{\bf b}^{\rm tr}\rVert^2\big((\gamma+1)\lVert({\bm C}^{\rm tr}+\widetilde\lambda I_d)^{-1}{\bf b}^{\rm tr}\rVert^2+1\big)\bigg],
\end{align}
and for IDG (${\bf b}^{\rm te}={\bf b}^{\rm tr},\,{\bm C}^{\rm te}={\bm C}^{\rm tr}$) we have
\begin{align}
    \Delta {\cal E}^A_{\rm IDG} =& 
    \frac{q\widetilde\lambda(1+\sigma^2)^2}{d}\bigg[(\lambda-\gamma)\lVert({\bm C}^{\rm tr}+\widetilde\lambda I_d)^{-1}{\bf b}^{\rm tr}\rVert^2 \nn
    & \qquad \qquad -\lambda\Big(2\widetilde\lambda({\bf b}^{\rm tr})^\top({\bm C}^{\rm tr}+\widetilde\lambda I_d)^{-3}{\bf b}^{\rm tr}
    + q\widetilde\lambda\lVert({\bm C}^{\rm tr}+\widetilde\lambda I_d)^{-1}{\bf b}^{\rm tr}\rVert^2+q\widetilde\lambda(1+\sigma^2)^2\lVert({\bm C}^{\rm tr}+\widetilde\lambda I_d)^{-1}{\bf b}^{\rm tr}\rVert^4\Big)\bigg].
\end{align}
\end{widetext}
One can verify that \cite{pehlevan25}
\begin{align}
    ({\bf b}^{\rm tr})^\top({\bm C}^{\rm tr}+\widetilde\lambda I_d)^{-1}{\bf b}^{\rm tr} \le 1 , \quad \lVert({\bm C}^{\rm tr}+\widetilde\lambda I_d)^{-1}{\bf b}^{\rm tr}\rVert^2 \le \frac{1}{\widetilde\lambda},
\end{align}
which also imply
\begin{align}
    q\le \frac{1}{\lambda}, \quad ({\bf b}^{\rm tr})^\top({\bm C}^{\rm tr}+\widetilde\lambda I_d)^{-3}{\bf b}^{\rm tr} \le \frac{1}{\widetilde\lambda^2}.
\end{align}
It follows that $\Delta{\cal E}^A_{\rm ICL}$ and $\Delta{\cal E}^A_{\rm IDG}$ can be bounded as
\begin{align}
    |\Delta{\cal E}^A_{\rm ICL}| \le& \frac{(1+\sigma^2)^2}{\lambda d}\left[4\gamma+2+(1+\sigma^2)^2\Big(1+\frac{2\gamma+1}{\lambda}\Big)\right], \nn
    |\Delta{\cal E}^A_{\rm IDG}| \le& \frac{(1+\sigma^2)^2}{\lambda d}\left[4\lambda+2\gamma+(1+\sigma^2)^2\right].
\end{align}
We conclude that, for any small but nonzero $\lambda$, all terms of the test errors containing ${\bf b}^{\rm tr}$ vanish in the joint asymptotic limit. This is consistent with the fact that, in \cite{pehlevan25}, the ridgeless limit ($\lambda=0$) was taken only \textit{after} the joint asymptotic limit.

\section{Derivation of the annealed ICL and IDG Errors}\label{app:derivation_of_annealed_errors}
 
In this appendix, we collect the spectral manipulations leading to the annealed ICL and IDG errors used in the main text. Note first that, since the eigenvalues $\{\lambda_j\}_{j=1}^d$ of ${\bm C}^{\rm tr}$ follow the MP distribution $\rho_{\rm MP}(\lambda)$ in the joint asymptotic limit, we have the asymptotic equivalence
\begin{align}
    \frac{1}{d}\sum_{j=1}^{d}\frac{1}{\lambda_j+z} &= \int {\rm d}\lambda'\frac{\rho_{\rm MP}(\lambda')}{\lambda'+z}=I_\kappa(z) , \nn
    \frac{1}{d}\sum_{j=1}^{d}\frac{1}{(\lambda_j+z)^2} &= -\frac{{\rm d}I_\kappa(z')}{{\rm d}z'}\bigg|_{z'=z} \equiv -I_\kappa'(z) , 
\end{align}
where $I_\kappa(z)$ is the Stieltjes transform of the MP distribution in Eq.~\eqref{Stieltjes-app}. This implies two spectral identities,
\begin{align}
    \frac{1}{d}\sum_{j=1}^d\frac{\lambda_j}{\lambda_j+\widetilde\lambda} 
    &=1-\widetilde\lambda I_\kappa(\widetilde\lambda), \nn
    \frac{1}{d}\sum_{j=1}^d\left(\frac{\lambda_j}{\lambda_j+\widetilde\lambda}\right)^2
    &=1-2\widetilde\lambda I_\kappa(\widetilde\lambda)
    -\widetilde\lambda^2 I_\kappa'(\widetilde\lambda) ,
    \label{spectral_identities_app}
\end{align}
which, substituted into the spectral representations of the annealed errors in Eq.~\eqref{E_annealed-spectral}, give Eq.~\eqref{E_annealed-Stieltjes}. For the ICL error, we first rewrite the spectral representation as 
\begin{align}
    {\cal E}^A_{\rm ICL}=\sigma^2+1-\frac{2}{d}\sum_{j=1}^d
    \frac{\lambda_j}{\lambda_j+\widetilde\lambda}+\frac{1+\gamma}{d}\sum_{j=1}^d\left(\frac{\lambda_j}{\lambda_j+\widetilde\lambda}
    \right)^2 , 
\end{align}
and then apply the identities in Eq.~\eqref{spectral_identities_app}.

To show the non-negativity of the difference ${\cal E}^A_{\rm ICL}-{\cal E}^A_{\rm IDG}$, we first obtain its expression from Eqs. \eqref{E_annealed-spectral} and \eqref{E_annealed-Stieltjes},
\begin{align}
    {\cal E}^A_{\rm ICL}-{\cal E}^A_{\rm IDG} =& \frac{\widetilde\lambda^2}{d} \sum_j \frac{1-\lambda_j}{(\lambda_j+\widetilde\lambda)^2} \nn
    =& -\widetilde\lambda^2[(1+\widetilde\lambda)I_\kappa'(\widetilde\lambda)+I_\kappa(\widetilde\lambda)].
\end{align}
Using
\begin{align}
    I_\kappa'(z) = -I_\kappa(z)\frac{1+I_\kappa(z)/\kappa}{\sqrt{(z+1-1/\kappa)^2+4z/\kappa}} ,
\end{align}
the error difference can be rewritten as
\begin{align}
    {\cal E}^A_{\rm ICL}-{\cal E}^A_{\rm IDG} = \frac{\widetilde\lambda^2}{2}I_\kappa^2(\widetilde\lambda)\frac{N(\kappa,\widetilde\lambda)}{\sqrt{(\widetilde\lambda+1-1/\kappa)^2+4\widetilde\lambda/\kappa}} ,
\end{align}
where
\begin{align}
    N(\kappa,\widetilde\lambda) =& \frac{1}{\kappa}\sqrt{\left(\widetilde\lambda+1-\frac{1}{\kappa}\right)^2+\frac{4\widetilde\lambda}{\kappa}} + \frac{3-\widetilde\lambda}{\kappa} - \frac{1}{\kappa^2} \nn
    =& \frac{1}{\kappa}\left[\sqrt{4\widetilde\lambda+\left(1-\widetilde\lambda-\frac{1}{\kappa}\right)^2}+1-\widetilde\lambda-\frac{1}{\kappa}\right]+\frac{2}{\kappa} \nn
    \ge& \frac{2}{\kappa}.
\end{align}
This leads to the lower bound
\begin{align}
    {\cal E}^A_{\rm ICL}-{\cal E}^A_{\rm IDG} \ge \frac{\widetilde\lambda^2I_\kappa^2(\widetilde\lambda)}{\sqrt{(\widetilde\lambda\kappa+\kappa-1)^2+4\widetilde\lambda\kappa}} .
\end{align}
The error difference is strictly positive for $\widetilde{\lambda}>0$. Its limit as $\widetilde{\lambda}\to 0^+$ depends on the task diversity; Employing the small-$z$ behavior of $I_\kappa(z)$ in Eq.~\eqref{eq:mp-asymptotics}, we obtain
\begin{align}
    \lim_{\widetilde{\lambda}\to0^+}\left({\cal E}^A_{\rm ICL}-{\cal E}^A_{\rm IDG}\right)=\max(1-\kappa,0).
\end{align}
Thus, for $\kappa<1$, the ICL--IDG error gap remains $1-\kappa$ because of the zero-eigenvalue sector of ${\bf C}^{\rm tr}$, whereas it vanishes for $\kappa\ge 1$.

To show Eq.~\eqref{zero-eigenspace-projector}, which reveals the origin of the nonanalyticity at $\kappa=1$, we use the spectral decomposition of ${\bm C}^{\rm tr}$ in Eq.~\eqref{Ctr-diag}:
\begin{align}
    \widetilde\lambda\left({\bm C}^{\rm tr}+\widetilde\lambda I_d\right)^{-1} = \sum_{j=1}^d\frac{\widetilde\lambda}{\lambda_j+\widetilde\lambda} {\bf m}_j{\bf m}_j^\top .
\end{align}
In the limit $\widetilde\lambda\to0$, the coefficients satisfy
\begin{align*}
    \frac{\widetilde\lambda}{\lambda_j+\widetilde\lambda}\longrightarrow\begin{cases}
        0, & \lambda_j>0, \\
        1, & \lambda_j=0,
    \end{cases}
\end{align*}
which implies Eq.~\eqref{zero-eigenspace-projector}. Eq.~\eqref{annealed-cusp} is then obtained by exploiting the properties of a projector, \textit{e.g.} ${\bm \Pi}_0^2={\bm \Pi}_0$.

\end{document}